\RequirePackage{fix-cm}
\documentclass[twocolumn]{svjour3}          
\smartqed  
\usepackage{graphicx}
\usepackage{natbib}
\usepackage{amsmath}
\usepackage[table]{xcolor}
\usepackage[dvipdfm]{hyperref}
\hypersetup{
   colorlinks=true,             
   linkcolor=blue,              
   urlcolor=blue,               
   anchorcolor=blue,            
   citecolor=blue,              
}

\definecolor{grey80}{rgb}{0.90,0.90,0.90}

\journalname{Celestial Mechanics and Dynamical Astronomy}
\begin{document}

\title{Optimization of Artificial Neural Networks models applied to
    the identification of images of asteroids' resonant arguments}

   \author{V. Carruba$^{1}$  \and
           S. Aljbaae$^{2}$  \and
           G. Carit\'{a}$^{2}$ \and
           R. C. Domingos$^{3}$ \and
           B. Martins$^{1}$ \and
   }
   
   \authorrunning{Carruba et al. 2021} 
      \institute{V. Carruba \at
     \email{\href{valerio.carruba@unesp.br}{valerio.carruba@unesp.br}} \\
     Orcid ID: 0000-0003-2786-0740 \\
      \and
      \begin{itemize}
        \item [$^{1}$] S\~{a}o Paulo State University (UNESP), School of Natural Sciences and Engineering, Guaratinguet\'{a}, SP, 12516-410, Brazil
        \item [$^{2}$] National Space Research Institute (INPE), Division of Space Mechanics and Control, C.P. 515, 12227-310, S\~{a}o Jos\'e dos Campos, SP, Brazil
        \item [$^{3}$] S\~{a}o Paulo State University (UNESP), Sao Jo\~{a}o da Boa Vista, SP, 13876-750, Brazil
      \end{itemize}
   }
   \date{Received: date / Accepted: date}   
   \maketitle

\begin{abstract}
  The asteroidal main belt is crossed by a web of mean-motion and
  secular resonances, that occur when there is a commensurability
  between fundamental frequencies of the asteroids and planets.
  Traditionally, these objects were identified by visual inspection of the
  time evolution of their resonant argument, which is a combination
  of orbital elements of the asteroid and the perturbing planet(s).
  Since the population of asteroids affected by these resonances is, in
  some cases, of the order of several thousand, this has become a taxing
  task for a human observer.  Recent works used Convolutional Neural
  Networks ({\it CNN}) models to perform such task automatically.  In this
  work, we compare the outcome of such models with those of some of
  the most advanced and publicly available {\it CNN} architectures,
  like the {\it VGG, Inception} and {\it ResNet}. The performance
  of such models is first tested and optimized for overfitting issues, using
  validation sets and a series of regularization techniques like
  data augmentation, dropout, and batch normalization.
  The three best-perfor\-ming models were then used to predict the labels
  of larger testing databases containing thousands of images.
  The {\it VGG} model, with and without regularizations, proved
  to be the most efficient method to predict labels of large datasets.
  Since the Vera C. Rubin observatory is likely to discover up to
  four million new asteroids in the next few years, the use
  of these models might become quite valuable to identify
  populations of resonant minor bodies.
  
  \end{abstract}

\section{Introduction}
\label{sec: intro}

Neuron networks in biological systems served as a model for artificial neural
networks ({\it ANNs}, hereafter). A typical $ANN$ shows a neural network
with an input, a hidden layer, and an output layer by having several
layers between the input and output layers.  There are many different sizes
and forms of neural networks, but they all share the same fundamental
building blocks: neurons, weights, biases, and activation functions.
These parts work similarly to the human brain and can be trained using the
same techniques as other traditional machine-learning techniques.
Among $ANN$, Convolutional Neural Networks ($CNN$) have proven to be
successful in challenging computer vision problems, achieving cutting-edge
outcomes on tasks like picture classification while also participating in
hybrid models for brand-new difficulties like object localization, image
captioning, and more. $CNN$s are regularized versions of multi-layer
perceptrons, which are fully connected networks, where each neuron in one
layer is coupled to every neuron in the layer above it.
These networks are susceptible to data overfitting due to their "full
connectivity." Regularization, or the avoidance of overfitting, can be
achieved in many ways, such as by punishing training-related
variables (such as weight reduction) or by lowering connectivity (
skipped connections, dropout, etc.). Advanced $CNN$ architectures,
like the VGG model \citep{2014arXiv1409.1556S}, the inception
architecture \citep{szegedy2015going}, and the Residual Network, or ResNet
\citet{he2015residual} have been among the most successful models
in recent years to solve multi-class problems of images classifications,
each winning awards at different editions of the ImageNet Large Scale
Visual Recognition Challenge, or ILSVRC.  Interested readers can find
more details on the history of the development of $CNN$s in
\citet{brownlee_2020}.

Applications of $CNN$ in astronomy, and in particular in the field of small
bodies dynamics, have been, so far, somewhat limited.  Examples of recent
applications include {\it DeepStreaks} \citet{2019MNRAS.486.4158D} and 
{\it Tails} \citet{2021AJ....161..218D}, which are convolutional neural
network, deep-learning systems designed to efficiently
identify streaking fast-moving near-Earth objects and comets
identified by the Zwicky Transient Facility (ZTF).
Additionally, asteroids can be automatically categorized into their
taxonomic spectrum classes using neural networks.
Using datasets imitating Legacy Survey of Space and Time (LSST)
and GAIA observations, \citet{10.3389/fspas.2022.816268, Pentilla_2021}
employed feed-forward neural networks to forecast
asteroid taxonomy with the Bus-DeMeo classification.  According to the
results, neural networks can reliably detect asteroids' taxonomic classes. 

\begin{figure}
  \centering
  \centering \includegraphics[width=3.5in]{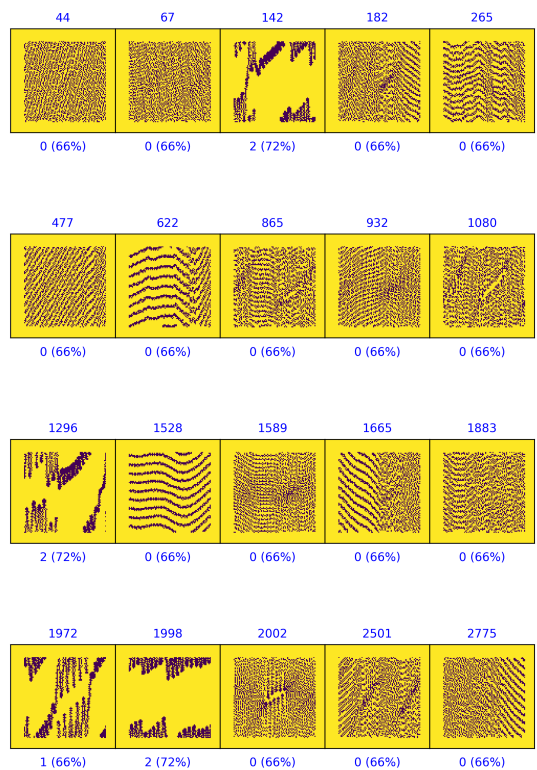}
  \caption{An application of the \citet{2021MNRAS.504..692C} $ANN$ model
    to a set of 20 images of resonant arguments of asteroids affected by
    the M1:2 mean-motion resonance with Mars. The number above each resonant
    argument is the asteroid identification, while the number
    at the bottom is their class, 0 for circulation,
    1 for switching orbits, and 2 for librating ones), with the
    associated probability of the asteroid belonging to that class,
    as computed by the $ANN$ model.}
\label{fig: M12_example}
\end{figure}

Regarding the main theme of this work, $ANN$s were first used for the
automatic classification of asteroids' resonant angles
in \citet{2021MNRAS.504..692C} for asteroids affected by the 1:2
external mean-motion resonance with Mars (M1:2 hereafter).  
For asteroids affected by this resonance, the resonant argument is:

\begin{equation}
  {\sigma}_1 = 2 \lambda - {\lambda}_M - {\varpi}_M,
  \label{eq: sigma_1}
\end{equation}

\noindent where $\lambda = M+ \Omega + \omega$ is the mean longitude,
$\varpi = \Omega + \omega$, with $\Omega$ being the longitude of
the node and $\omega$ the argument of pericenter, and where the suffix
$M$ identifies the planet Mars, can have three kinds of behavior:
circulation, libration, and switching orbits.  For circulation
orbits the resonant argument will cover all possible values
from $0^{\circ}$ to $360^{\circ}$.  For librating orbits,
the resonant argument will oscillates around a stable equilibrium
point, while for switching orbits the resonant argument will
alternate phases of libration and circulation.  An example of
the application of \citet{2021MNRAS.504..692C} $ANN$ model is
shown in figure~(\ref{fig: M12_example}), where the probability of
each resonant argument image to belong to a given class is shown at the bottom
of each panel for 20 asteroids.  Asteroid 44 Nysa is in a circulating
orbit, 1972 Yi Xing is in a switching orbit, and 1998 Titius is
on a librating orbit around the $0^{\circ}$ equilibrium point.

More recently, a similar model was used by \citet{2022MNRAS.514.4803C}
to classify the orbits of asteroids affected by the ${\nu}_6$
secular resonance.  The main difference with the
\citet{2021MNRAS.504..692C} model was the presence of an
additional class, since for ${\nu}_6$ asteroids libration
can occur around two equilibrium points, $0^{\circ}$ (aligned
libration) and $180^{\circ}$ (antialigned libration).

The main goal of this paper is to investigate how the use of
more advanced model architectures can improve the classification
of asteroids dynamical regimes, and how employing regularization
techniques can reduce the problem of the overfitting of the training
data-set.  For these purposes, $CNN$s models
like the VGG, Inception, and ResNet, with and without regularization
techniques, will be applied to publically available resonant arguments
images databases, like those for the M1:2 and ${\nu}_6$ resonance.
Selecting the most efficient method to fit these databases, for small
and large testing data-sets, will be the ultimate goal of this work.

This paper is so divided: in section~(\ref{sec: CNN_intro}) we revise
some basic properties of $CNN$. In section ~(\ref{sec: ANN_models})
we revise the architecture of some of the recently published
more advanced $CNN$ models.  Methods to evaluate the models' performances
will be discussed in section~(\ref{sec: over_under_fit}),
while in section~(\ref{sec: perf_impr}) we will discuss approaches
to improve the models' performances through regularization.
In section~(\ref{sec: appl}) we will apply $CNN$ models, regularized
or not, to the two asteroids' resonant arguments images databases to
select the three best performing models for each dataset.
In section~(\ref{sec: larger_data}) these models will be tested
with large testing sets, to ultimately select the best performing
method for each set. Finally, in section~(\ref{sec: concl}) we will
discuss our conclusions.

\section{Convolutional Neural Networks, some basic concepts}
\label{sec: CNN_intro}

Convolutional neural networks, or $CNNs$, are a type of neural network models
originally designed to work with two-dimensional picture data.
Their name derives from the convolutional layer, which is an
important layer of the model. Convolution is a linear procedure involving
the multiplication of a two-dimensional array of weights, often
referred to as a filter, with an input array.
This is achieved by  multiplying the input and filter's
filter-sized patch element-by-element before summing the results, which
produces a single value.  This operation is occasionally referred
to as the "scalar product", because it produces a single integer.
Since the same filter might be multiplied by
the input array several times at various points on the input, it
is better to use a filter that is smaller than the input.
The filter is applied sequentially from left to right and top to bottom
to each overlapped piece of the input data that has the same size of the
filter. 

This idea of applying the same filter to an image in a systematic way is a
strong concept. If a filter is designed to identify a particular type of
input feature, applying the filter methodically across the entire image
provides the possibility to find the feature anywhere in the data.
Translation invariance, which is the name of this quality, denotes the general
interest in whether a feature exists rather than its location. 

The filter is multiplied once with a filter-sized piece of the input
array to create a single value. When the filter is applied uniformly
throughout the input array, the result is a two-dimensional array of
output values that represent the
input filtering. As a result, the two-dimensional
output array of this procedure is referred to as a feature map. Each
value in a feature map can be processed through a nonlinearity, such as a
Rectified Linear Unit (ReLU), once the feature map has been built.
If the input value is negative,
the ReLU activation function is intended to be 0. For positive or zero
values, it is intended to be the input value itself, $y = x$. 

The disadvantage of convolutional layers' feature map output is that
it captures the precise location of features in the input.
Even little adjustments to the position of
the feature in the input image will produce a different feature map.
A common solution to this issue is to utilize a pooling layer, usually
added after the convolutional layer. In some models, this pattern may
be repeated once or more.  The pooling layer builds new pooled
feature maps with the same amount of features by working separately on
each feature map. In that it requires choosing a pooling procedure,
pooling is comparable to applying a filter to feature maps.
The pooling operation or filter, for instance, is often applied
with a stride of 2 pixels and is typically smaller than the feature map. 

This implies that each feature map will always be compressed
by a factor of two by the pooling layer, i.e., each dimension will
be halved, producing a feature map with a quarter as many pixels or values. 
A pooling layer applied to a 6 $\times$ 6 (36 pixels) feature map, for
example, will provide a map of 3 $\times$ 3 pixels
(9 pixels). Rather than being learned, the pooling operation is defined.
Two common functions used for pooling operations are the
{\bf average pooling}, which computes the average value for each considered
patch on the feature map, and the {\bf maximum pooling}, where the maximum
value is instead used.
The result of using a pooling layer and creating downsampled or pooled
feature maps is a summary of the features found in the input.
Their advantage resides in the fact that even minute adjustments in the
location of the feature in the input that the convolutional layer picks
up will result in the feature remaining in the same location on the pooled
feature map. The local translation invariance of the model is a
benefit added by the pooling layer.

\section{$ANN$ models}
\label{sec: ANN_models}

The first successful model for image classification was
LeNet-5, outlined in the paper by
\citet{lecun-gradientbased-learning-applied-1998}.
The system was created to solve a handwritten character recognition
problem, and it was tested using the MNIST standard dataset, attaining a
classification accuracy of about 99.2\%. (or a 0.8 percent error rate).
This model was subsequently characterized as the core component of a
larger system known as Graph Transformer Networks. 

Advancements in the field of image recognition came later on,
also because of the efforts of the ImageNet project and its sponsored
computer vision competition, the ImageNet Large Scale Visual Recognition
Challenge (ILSVRC). The ImageNet project is a large visual
database created for use in visual object recognition software research. 
The project has several million of images that have been hand-annotated
to indicate what objects are depicted.  Bounding boxes are provided
for at least one million of the images.  ILSVRC is an annual software
contest, that has been held since 2010, where software programs
compete to accurately identify and detect objects and sceneries.
AlexNet refers to the model developed by \citet{NIPS2012_c399862d} to
compete in the ILSVRC-2010 competition for the classification of
objects photographs into 1000 different categories. Before the
creation of AlexNet, this task was deemed to be too difficult for the
then-current capacities of computer vision approaches.  AlexNet success
sparked a competition that inspired many more subsequent new models,
several of which were applied to the same ILSVRC task of the original
AlexNet problem.

\begin{figure}
  \centering
  \centering \includegraphics[width=3.5in]{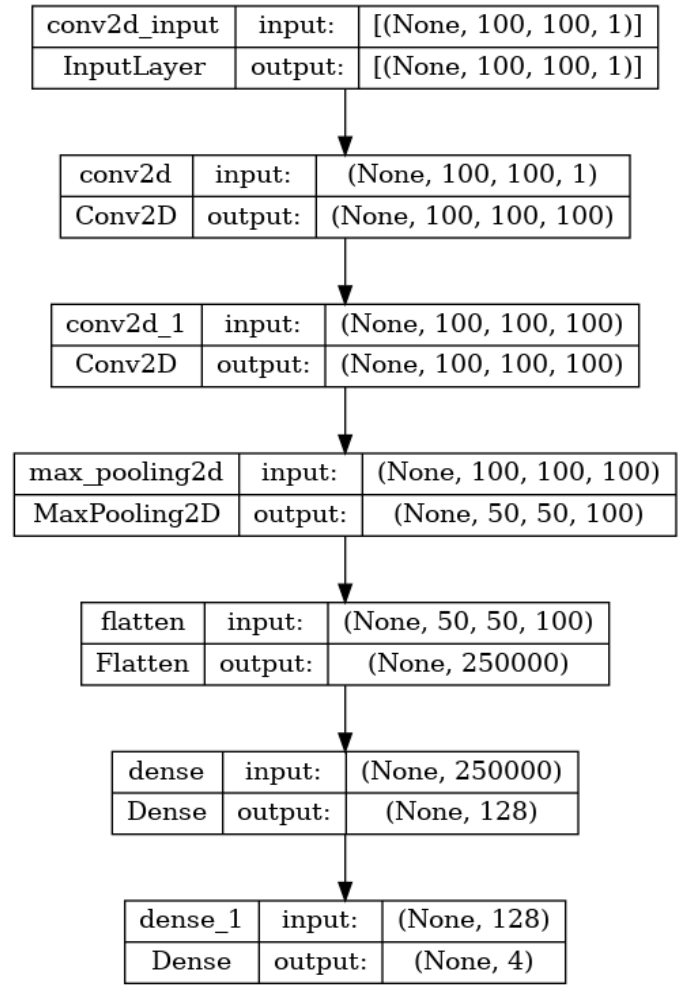}
  \caption{The architecture of the VGG model, as implemented in
    this work.  The initial input dimension of [(None, 100, 100, 1)]
    refers to the dimension of the images of 100 $\times$ 100 pixels, with
    just one color channel (1), associated with a black and white picture.
    This input is fed to the first convolutional layer, that processes it and
    feeds its output to the next layer, until reaching the final layer
    with a dimension of 4, associated with the four possible classes
    studied in this problem, i.e., the classification of images
    of resonant arguments of asteroids interacting with the ${\nu}_6$
    secular resonance.}
\label{fig: arch_VGG}
\end{figure}

\citet{2014arXiv1409.1556S} paper aimed at standardizing
architecture designs for convolutional networks,
while also developing better performing models.
Their architecture is commonly referred to as VGG, after their lab's name,
the Visual Geometry Group at Oxford, and was developed and tested
for the ILSVRC-2014 competition. Their use of several small filters
has become a new standard. Most, but not all, convolutional layers are
followed by max pooling layers.
Another notable characteristic is the large number of filters, which
rises with model depth, starting with 64 and proceeding through
128, 256, and 512 filters at the model's feature extraction end.
Among many possible architecture versions of the VGG model, two
are most commonly used: the VGG-16 and VGG-19,
which are named after the number of learned layers, 16 and 19, respectively.
These architectures showed the best performance and depth among several
VGG versions that were built and assessed.
Finally, the VGG study was among the first to make 
model weights freely available, and this set a trend among
researchers in the area of computer vision. As a result, in transfer learning,
pre-trained models such as VGG are routinely employed as a starting point.
The architecture of the VGG model, as defined in \citet{brownlee_2020}
and applied in this work, is shown in figure~(\ref{fig: arch_VGG}).
Interested readers can find more details on the architecture and reasons
for the choice of free parameters and number of layers of the
models used in this work in \citet{brownlee_2020}.
  
\begin{figure}
  \centering
  \centering \includegraphics[width=3.5in]{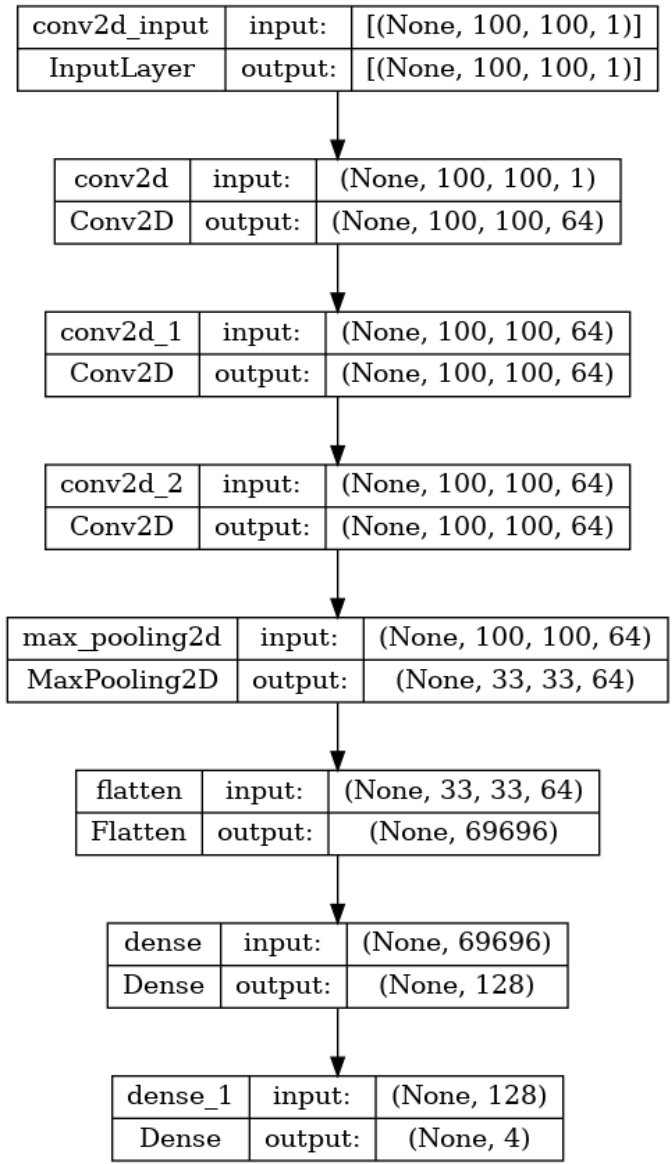}
  \caption{The architecture of the Inception model, as implemented in
    this work. Note the use of a stack of convolutional layers at
    the beginning of the architecture, typical of this model.  See the
    text for more details on this architecture.}
\label{fig: arch_inception}
\end{figure}

In their 2015 study titled Going Deeper with Convolutions,
\citet{szegedy2015going} offered significant advancements in the
application of convolutional layers. The authors of the paper suggested an
architecture called inception
and a particular network named GoogLeNet that came in first
place in the ILSVRC challenge's 2014 round.  The most important innovation
behind the inception models, and the main reason for their name, is
the inception module. This is a stack of parallel convolutional layers with
various filter sizes (such as 1 $\times$ 1, 3$\times$ 3, and 5 $\times$ 5), as
well as a pooling layer with a 3 $\times$ 3 maximum, the output of which is
subsequently concatenated.  A problem with implementations of this model is
that the filters' number (depth or channels) starts to quickly increase in a
naive implementation of the inception model, especially when inception modules
are stacked. It can be computationally expensive to perform convolutions
on many filters using larger filter sizes (like 3 and 5).
To remedy this, 1 $\times$ 1 convolutional layers are employed before the
3 $\times$ 3 and 5 $\times$ 5 convolutional layers, as well as after the
pooling layer, to reduce the number of filters in the inception model.
The architecture of the Inception model, as used in this work, is shown
in figure~(\ref{fig: arch_inception}).

\begin{figure}
  \centering
  \centering \includegraphics[width=3.5in]{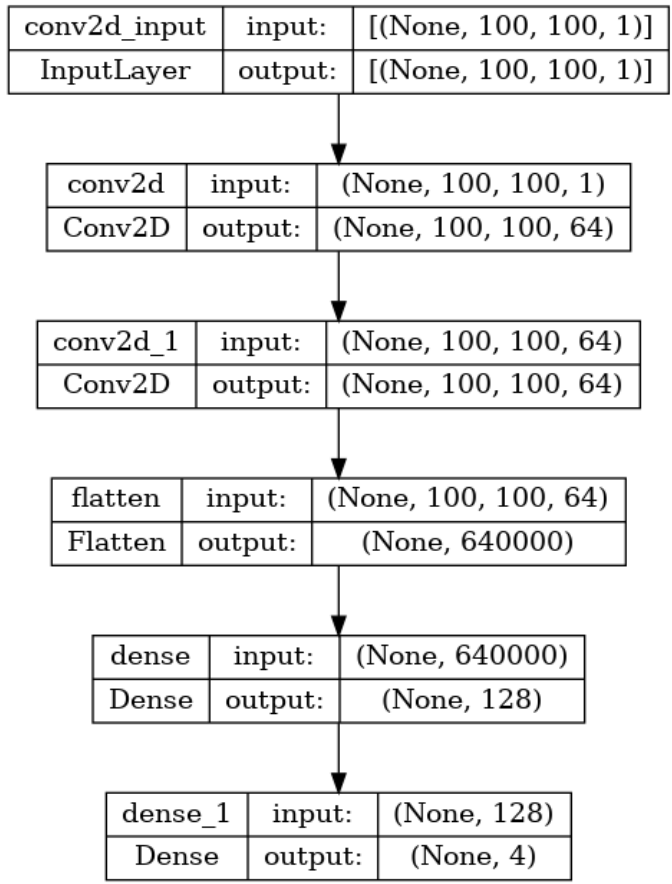}
  \caption{The architecture of the ResNet model, as implemented in
    this work. Note the presence of residual blocks, that are
    patterns comprising two convolutional layers activated by ReLU.
    See the text for more details on this architecture.}
  \label{fig: arch_resnet}
\end{figure}

A Residual Network, or ResNet, was introduced by \citet{he2015residual} 
as an extremely deep model that was successful
in the 2015 ILSVRC challenge. Impressively, their model has 152 layers.
The concept of residual blocks that utilize shortcut connections is crucial
to the architecture of the model. These are just connections
where the input is transmitted to a deeper layer while
remaining unchanged (i.e. skipping the following layer).
A residual block is a pattern comprising two convolutional layers
activated by ReLU, where the output and input of the block are combined. 
This is called the shortcut connection. If the input shape to
the block is different from the form of the output of the block,
a projected version of the input is employed. 
The model starts with what the authors define as a "plain network".
Quoting \citet{brownlee_2020}, ``this is a deep convolutional neural
network inspired by the VGG with small
filters, grouped convolutional layers that are followed by
no-pooling layers, and an average pooling after the feature
detector portion of the model, before an output layer with a
softmax activation function''. A residual network that defines residual
blocks is created by introducing shortcut connections to the
plain network. The shortcut connection's input often has the same shape
as the residual block's output. The architecture of the ResNet model is shown
in figure~(\ref{fig: arch_resnet}).

The ResNet model was the last model revised in this work.
In the next section, we are going to introduce methods to evaluate
the $CNN$ models' performance.

\section{Evaluating the models' performance}
\label{sec: over_under_fit}

In order to evaluate the performance of deep learning models, two metrics
are often used, {\bf {\textit Cross-Entropy loss} \citep{Cox_1958} and
{\textit Accuracy} \citep{Metz_1978}}.
Loss/cost functions are employed to optimize the model during training
while working on a machine learning or deep learning problem.
Almost generally, the goal is to reduce the loss function.
The model performs better if the loss is lower.
One of the most significant cost functions is the Cross-Entropy loss.
To understand the cross entropy loss one has first to grasp the concept
of the softmax activation function.   The Softmax activation function is
typically positioned as the deep learning model's final layer. 
The activation function known as Softmax scales numbers and logits
into probabilities. Softmax produces a vector (let's say $\vec{v}$) with a
probability for each potential result. For all conceivable outcomes or
classes, the probabilities in vector $\vec{v}$  add up to one.
The mathematical definition of Softmax is 

\begin{equation}
S(y)_i = \frac{exp(y_i)}{\sum_{j=1}^{n} exp(y_j)},
\label{eq: softmax}
\end{equation}

\noindent where $y$ is an input vector consisting of $n$ elements for
$n$ classes. Consider a CNN model that attempts to categorize a picture as
either a dog, cat, horse, or cheetah (four possible results/classes).
A vector of logits, L, is produced by the final (completely connected)
layer of the CNN and is then sent through a Softmax layer, which converts
the logits into probabilities, P. For each of the 4 classes, these
probabilities represent the model predictions. 

For example, let us assume that the model will classify the picture
of a dog using the Softmax function with a probability 0.775 of being a dog,
0.116 for being a cat, 0.039 for being a horse, and 0.070 of being a cheetah.  
The goal of a deep learning model is to make the output as close as feasible
to the desired result.  The Cross-Entropy loss function, also called
logarithmic loss, log loss or logistic loss,
is defined in terms of the actual class desired output 0 or 1.
A score/loss is calculated by penalizing the probability based on how different
it is from the actual expected value. Because of the penalty's logarithmic
structure, significant differences close to 1 receive a large score, while
minor differences close to 0 receive a small score. 
Mathematically, the Cross-Entropy loss function is defined as 

\begin{equation}
  L_{CE} = - \sum_{i=1}^{n} t_i log(p_i),
\label{eq: cross_entropy_loss}
\end{equation}

\noindent where $t_i$ is the truth label and $p_i$ is the probability
for the $i-th$ class.  For the case of the classification of the dog's image,
the Cross-Entropy loss function will be minimal for the case of the image
being labeled as a dog, larger in the other cases, and will reach a maximum
for the horse class.  For the cases where we have multiple classes
classification, either where the labels are integers, or one-hot-encoded,
\textit{sparse categorical cross-entropy} and
\textit{categorical cross-entropy} can be defined.   Interested readers
are referred to the Keras API reference (https://keras.io/api/,
\citep{Chollet_2018}).

One parameter for assessing classification models is accuracy.
The percentage of predictions one's model correctly predicted is
known as Accuracy. 
Formally, Accuracy has the following definition:

\begin{equation}
Accuracy = \frac{Number~of~correct~predictions}{Total~number~of~predictions}.
\label{eq: accuracy_raw}
\end{equation}

\noindent
Accuracy can also be determined in terms of positives and negatives for
binary classification, as seen below: 

\begin{equation}
Accuracy = \frac{TP+TN}{TP+TN+FP+FN},
\label{eq: accuracy}
\end{equation}

\noindent where $TP$ = True Positives, $TN$ = True Negatives, $FP$ = False
Positives, and $FN$ = False Negatives. For the case of multiple class
classification, where each class is identified by an integer label, a
{\textit SparseCategoricalAccuracy} can be defined.  Interested readers
can find more information in the Keras API reference (https://keras.io/api/,
\citep{Chollet_2018}).

One of the most common limitations of deep learning models is their tendency
to overfit the training data.   Overfitting occurs when the neural network
model learns all the minute features of the training test.  While this
may result in nominally excellent values of Cross-Entropy loss and
Accuracy, the ability of the model to make reliable predictions
on test sets different than the training one is somewhat limited.
Overfitting may occur for models with a very large number of trainable
parameters, or for models that trained for too long.  Conversely,
underfitting may occur when the train data can still be improved upon.
This may occur for a variety of reasons, including inadequate model
strength, excessive regularization, or insufficient training time.
This indicates that the network has not picked up any useful patterns from
the training data. With the latest generation of $ANN$ models, overfitting
tends to be a more common occurrence than underfitting.

\begin{figure}
  \centering
  \centering \includegraphics[width=3.5in]{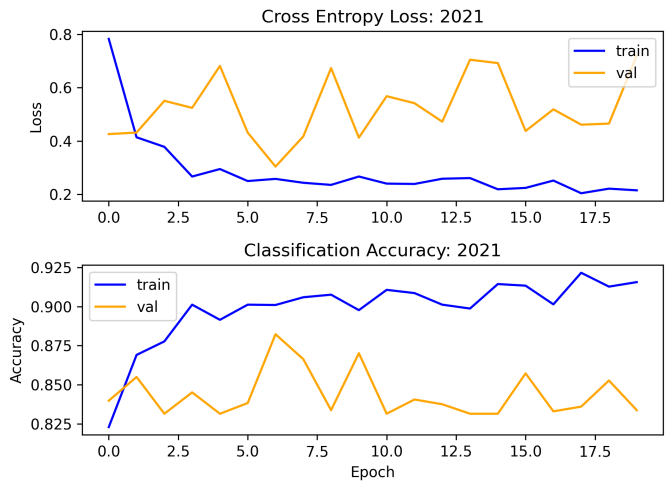}
  \caption{Cross entropy loss (top panel) and classification accuracy
    (bottom panel) as a function of epoch for an application of the
    \citet{2022MNRAS.514.4803C} $ANN$ model.  The distance between
    the training and validation metrics, and the fact that the accuracy
    for the training metric is consistently higher suggests that this
    model may be affected by overfitting.}
\label{fig: CEL+ACC_2021}
\end{figure}

A method to test for overfitting is to test the model on a validation
set.  A validation set is a part of the database over which the model
trained on the training set can be tested.  Usually, it is made of
about 20\% of the total database.  If the model is overfitting the
training set, while values of the training Cross-Entropy loss
will go down and Accuracy will increase and be close to 1, the same
will not be observed for these metrics applied to the validation
set. Figure~(\ref{fig: CEL+ACC_2021}) shows a typical case where
overfitting is observed.  The actual performance of the model can be
improved using methods such as Data Augmentation, Dropout, and Batch
Normalization, which will be discussed in the next section.
Generally speaking, the lower the difference between the values
of Cross-Entropy loss and Accuracy of the training and validation
sets, the better the actual performance of the model.

\section{Improving the performance: regularization methods}
\label{sec: perf_impr}

There are various methods available to improve a CNN algorithm's performance
and reduce overfitting.  Here we will focus on {\textit data augmentation,
dropout}, and {\textit batch normalization}.

\begin{figure}
  \centering
  \centering \includegraphics[width=3.5in]{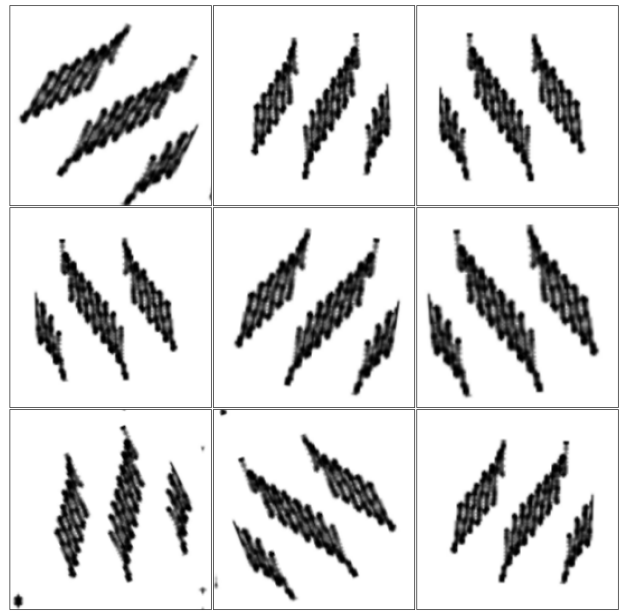}
  \caption{An example of Data Augmentation techniques applied to an
    image in the ${\nu}_6$ database.  We applied three instances each of a
    random flip, random rotation and random zoom.}
\label{fig: data_augmentation}
\end{figure}

{\textit Data augmentation} entails creating duplicates of the training
dataset's samples with minor random changes. This has a regularizing effect by
allowing the model to learn the same broad features but in a more
generic way while also expanding the training dataset.
A variety of data augmentation techniques could be used.
Given that the datasets that we are going to investigate consist of
tiny images of resonant arguments, we will not use augmentation that
severely distorts the images in order to preserve and utilize relevant
information. The kinds of random augmentations that might be helpful include
flipping an image horizontally, making modest image shifts, and possibly
making minor zooming or cropping adjustments.
Figure~(\ref{fig: data_augmentation}) displays an example of data
augmentation obtained with the {\textit ImageDataGenerator} class in Keras for
an image from the ${\nu}_6$ database (see section~(\ref{sec: appl})
for more details on the use and applications of this dataset).
Our example includes basic augmentations, such as flipping the image
horizontally and shifting its height and breadth by 10\%.

{\textit Dropout} is a straightforward method that will erratically remove
nodes from the network. As the remaining nodes must adjust to fill in the
gaps left by the eliminated nodes, it has a regularizing effect.
By including new dropout layers to the model, where the number of nodes
deleted is controlled by a parameter, dropout can be added.
In terms of where to add the layers and how much dropout to utilize,
there are numerous patterns for adding dropouts to a model. In our case, we
will use a fixed dropout rate of 20\%, and add dropout layers after each
max pooling layer and after the fully connected layer.

Finally, another method to reduce overfitting is batch normalization.
\citet{brownlee_2020} defines this method as ``a technique for training
very deep neural networks that standardizes the inputs to a layer for each
mini-batch. This has the effect of stabilizing the learning process and
dramatically reducing the number of training epochs required to train deep
networks.''

Batch normalization employs a transformation that keeps the output mean and
standard deviation close to 0 and to 1, respectively.  Interested
readers can find more information on the method in \citet{brownlee_2020},
here we will use batch normalization as implemented in Keras
(https://keras.io/api/, \\ \citet{Chollet_2018}).

\section{Applications to images databases}
\label{sec: appl}

\begin{figure}
  \centering
  \centering \includegraphics[height=3.5in]{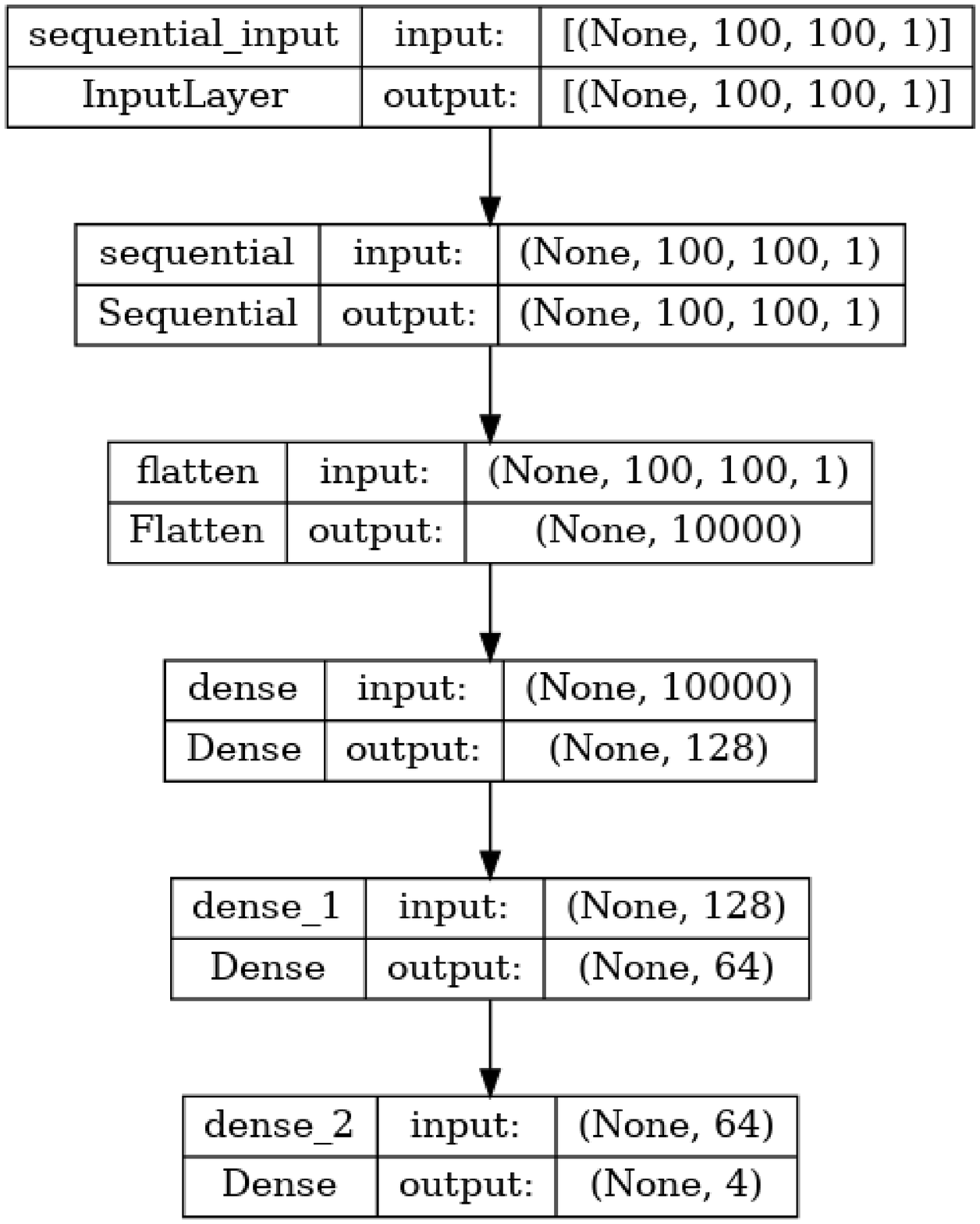}
  \caption{The architecture of the \citet{2022MNRAS.514.4803C} $ANN$
       model.}
\label{fig: 2021_architecture}
\end{figure}

The use of $ANN$ for the purpose of classification of images of
asteroid resonant arguments started in 2021 with the work
of \citet{2021MNRAS.504..692C}, where images of the resonant
arguments of 6440 asteroids
affected by the M1:2 exterior mean-motion resonance with Mars
were classified, and the dynamical regimes identified within
three categories (libration, circulation,
and orbits switching from one mode to the other).
Figure~(\ref{fig: 2021_architecture}) displays the architecture
of the 2021 $ANN$ model.

More recently, the same model was used by \citet{2022MNRAS.514.4803C}
to classify the orbits of more than 10000 asteroids affected by
the ${\nu}_6$ secular resonance into four classes, circulation,
aligned libration, where the resonant argument oscillates around
$0^{\circ}$, antialigned libration, with oscillations around $180^{\circ}$,
and switching behavior.  Both databases of asteroids' resonant arguments
are publically available on the authors' research group server.

The model proposed in 2021 allowed to speed up the classification
of asteroidal images, but it has some limitations.  Considering
the ${\nu}_6$ database, if we set aside a sample of 4875 asteroids images
of numbered asteroids as a training set, and a sample of 1350
numbered asteroids with identifications between 523584 and 607011,
we obtained Cross entropy loss and classification accuracy as a function 
of the epoch, for 20 epochs. Figure~(\ref{fig: CEL+ACC_2021}) shows our
results, that are clearly affected by overfitting.

In this work, we will explore how the use of three more advanced models,
the VGG, Inception, and ResNet architecture (see
figures~(\ref{fig: arch_VGG}, \ref{fig: arch_inception},
\ref{fig: arch_resnet}) for a summary of their
architectures, as used in this work), improved by the regularization
techniques seen in section~(\ref{sec: perf_impr}), can boost
the performance of image recognition for these kinds of problems.

To quantitatively compare the outcome of the different models
tested in this work, we also computed the elapsed time needed to
fit all models over 20 epochs, and the mean and maximum memory
allocation during this process. All simulations ran alone on a Xubuntu
workstation with an Intel I9-10900 processor with a DDR4 16GB 3200Mhz
net core memory.  We will start our analysis by studying the ${\nu}_6$
database.

\subsection{Asteroids affected by the ${\nu}_6$ resonance}
\label{sec: nu_6}

As the first step in our analysis, we will use the ${\nu}_6$ asteroid
database with the training, test, and validation structure described
at the end of section~(\ref{sec: appl}).
The first model applied to
this dataset was the VGG, and was implemented alone
and i) with data augmentation (DA), ii) data augmentation and dropout (DA+DO),
and, finally, iii) also including batch normalization (DA+DO+BN).
Our results are shown in figure~(\ref{fig: Metrics_VGG}), in the Appendix.

The results of the VGG model alone are clearly affected by overfitting.
DA improves the outcome of the Cross entropy loss function, but accuracy
is still not optimal.  Better results are obtained when both DA and DO
are applied, and they are not much improved if we also use BN.

For the case of the Inception model (see
figure~(\ref{fig: Metrics Inception}), in the appendix),
again, we observe overfitting
for the model alone without regularizations.  The situation improves
significantly for models with DA and DO, and there is almost no overfitting
if we also consider BN.  However, in this case, values of Accuracy are
significantly lower than for what was observed for the DA+DO case.

Finally, for the ResNet model (see figure~(\ref{fig: Metrics ResNet}), in the
appendix) we actually observe underfitting for the model alone.  The best
results are again obtained for the model with DA+DO.

\begin{table*}
  \begin{center}
    \caption{Execution time, memory and maximum memory allocation
      for the models tested in this work for the ${\nu}_6$ database.}
    \label{Table: time_memory_nu6}
    \begin{tabular}{|c|c|c|c|}
\hline
Model & Time    & Memory         & Max. Memory \\
      & [h:m:s] & alloc. [Bytes] & alloc. [Bytes] \\
\hline
VGG          &  0:25:52.6 &   3633632 & 194524205 \\
VGG+DA	     &	0:25:14.1 &   4951169 & 194528646 \\
VGG+DA+DO    &	0:11:41.7 &   5157584 & 194526710 \\
VGG+DA+DO+BN &	0:13:02.8 &   5185155 & 194527529 \\
\hline
Inception          & 0:26:27.0 &  5746597 & 194896854 \\
Inception+DA       & 0:27:47.7 &  5883557 & 194924103 \\
Inception+DA+DO    & 0:28:27.4 &  5863357 & 194940207 \\
Inception+DA+DO+BN & 0:29:29.2 &  6388938 & 195023989 \\
\hline
ResNet          & 0:17:13.1 &   4142248 & 331982732 \\
ResNet+DA       & 0:16:50.2 &   5735049 & 333427831 \\
ResNet+DA+DO    & 0:19:43.4 &   5655480 & 333496679 \\
ResNet+DA+DO+BN & 0:21:30.1 &   5824092 & 333666545 \\
\hline
\end{tabular}
\end{center}
\end{table*}

Our results in terms of execution time and memory allocation
are reported in the table~(\ref{Table: time_memory_nu6}).  The three best
performing models in terms of avoidance of overfitting, execution
time, and memory allocation were the VGG, ResNet, and Inception regularized
with Data Augmentation and Dropout. For the case of the Inception model,
better results in terms of overfitting were obtained for the
case that also includes Batch Normalization, but values of
accuracy were better for the model with just DA and DO.

In the next subsection, we will analyze the case of the M1:2 resonance database.

\subsection{Asteroids affected by the M1:2 resonance}
\label{sec: M12}

The \citet{2021MNRAS.504..692C} database
contains images of 5700 asteroids' resonant
arguments.  This dataset was divided into 4560 images for training (80\%
of the total) and 1140 images for validation (20\% of the total), with
100 images used for the test set.  The same models discussed
in section~(\ref{sec: nu_6}) were applied to this database, with
the only difference that the last Dense layer had 3 nodes instead
of 4, since there are only three possible classes for this set.
Our results for the VGG, Inception, and ResNet models, without and
with the three regularizations (DA, DO, BN) are shown in the
appendix, in figures~(\ref{fig: Metrics_VGG_M12},
\ref{fig: Metrics Inception_M12}, \ref{fig: Metrics ResNet_M12}).
Table~(\ref{Table: time_memory_m12})
displays our results in terms of execution time and memory allocation.

The best performing model in terms of overfitting avoidance
was the Inception, regularized with Data Augmentation and Dropout.
Surprisingly, the second-best performance was obtained by
the VGG model with no regularization.   Finally,
the third best model was ResNet, regularized with DA, DO, and BN.
In the next section, we will review how these models perform for
larger test databases.

\begin{table*}
  \begin{center}
    \caption{Execution time, memory and maximum memory allocation
      for the models tested in this work for the M1:2 database.}
    \label{Table: time_memory_m12}
    \begin{tabular}{|c|c|c|c|}
\hline
Model & Time    & Memory         & Max. Memory \\
      & [h:m:s] & alloc. [Bytes] & alloc. [Bytes] \\
\hline
VGG          &  0:24:32.0 &  4744976 &  183469190 \\
VGG+DA	     &	0:24:16.0 &  5186968 &  196222507 \\
VGG+DA+DO    &	0:25:22.7 &  5947897 &  194526710 \\
VGG+DA+DO+BN &	0:28:37.3 &  5368555 &  183121859 \\
\hline
Inception          & 0:24:55.8 & 4725164 & 183486485 \\
Inception+DA       & 0:25:18.7 & 6094216 & 183512476 \\
Inception+DA+DO    & 0:25:25.3 & 6311918 & 183519955 \\
Inception+DA+DO+BN & 0:27:18.3 & 6467734 & 183605621 \\
\hline
ResNet          & 0:15:53.6 & 4653649 & 332345991 \\
ResNet+DA       & 0:16:19.3 & 5153001 & 379370007 \\
ResNet+DA+DO    & 0:18:40.8 & 6006420 & 333848121 \\
ResNet+DA+DO+BN & 0:19:35.9 & 6178250 & 334018955 \\
\hline
\end{tabular}
\end{center}
\end{table*}

\section{Applications to larger test datasets}
\label{sec: larger_data}

In the previous section, we trained our models with large testing
and validation sets, but with small testing ones.  Here, the three
best-performing models selected in the previous section are tested
on larger databases for both the ${\nu}_6$ and M1:2 resonances.
For the case of the ${\nu}_6$ secular resonance, we consider a
set of 3000 asteroid resonant arguments' images obtained by
\citet{2022MNRAS.514.4803C} for multi-opposition asteroids affected
by this resonance.  For the M1:2 resonance, we selected 650
numbered objects with identifications larger than 523584 and lower
than 606158 that were not previously considered by \citet{2021MNRAS.504..692C}.

To assess the performance of the models, we used the accuracy, as defined
in equations~(\ref{eq: accuracy_raw}) and (\ref{eq: accuracy}), and the
F-beta score.  The F-beta score is a metric, commonly used for binary
classification problems, which is based on precision and recall.
The percentage of accurate predictions for the positive class is
calculated using the {\it precision} metric, which is given by:

\begin{equation}
precision = \frac{TP}{TP + FP}.
\label{eq: precision}
\end{equation}

\noindent 
{\it Recall}, defined as:

\begin{equation}
recall = \frac{TP}{TP + FN}, 
\label{eq: recall}
\end{equation}

\noindent 
determines the proportion of
accurate positive predictions for the positive class out of all possible
positive predictions.  While maximizing {\it recall} will reduce false-negative
errors, maximizing {\it precision} will reduce false-positive mistakes.

The harmonic mean of {\it precision} and {\it recall} is used to
calculate the {\it F-measure},  giving each the same weight. This makes it
possible to compare models and describe a model's performance while also
accounting for both {\it precision} and {\it recall} in a single score.
A generalization of the {\it F-measure}, called
the {\it F-beta-measure} includes a beta configuration parameter.
For the {\it F-measure} this is equal to 1.0, which serves as the default beta
value. In the calculation of the score, a smaller beta value, such as 0.5,
gives more weight to {\it recall} and less to {\it precision}, whereas a
greater beta value, such as 2.0, gives less weight to {\it recall} and
more weight to {\it precision}.  Following \citet{brownlee_2020}, we can define
{\it F-beta} as:

\begin{equation}
  {F-beta} = (1+beta^2) \times \frac{ precision \times recall} {beta^2 \times
    precision + recall}
  \label{eq: F-beta}
\end{equation}

\noindent
Interested readers can find more information on the {\it F-beta} score in
\citet{brownlee_2020}, chapter 22.   Here, we are using the {\it F-beta}
score as introduced by \citet{scikit-learn}, and as implemented 
in the scikit-learn API \citet{sklearn_api}, with a beta value of 0.5,
to give more weight to {\it recall}. Also, since we are dealing with
a multi-class imbalanced problem, we used the 'weighted' option
of the average parameter, which will calculate metrics for each
label and find their average weighted by their support, i.e., the
number of true instances for each label.  For more details on this
last procedure please see \citet{scikit-learn}.

Our results for the two datasets are shown in
table~(\ref{Table: ext_databases}).  We included in our runs
two versions of the VGG model, with and without regularizations,
since the unregularized VGG model was one of the best performing
for the M1:2 database, and we wanted to check its behavior for
both cases. The best performing
models for the ${\nu}_6$ and M1:2 resonances were the VGG, with and without
regularization, with the Inception models performing slightly worse,
followed by the ResNet models as a distant third.

\begin{table}
  \begin{center}
    \caption{Values of {\it accuracy} and {\it F-beta} for the
      best performing models selected in section~(\ref{sec: nu_6}) and
      (\ref{sec: M12}).}
    \label{Table: ext_databases}
    \begin{tabular}{|c|c|c|c|}
\hline
Database & Model & {\it Accuracy} & {\it F-beta} \\ 
\hline
${\nu}_6$ & VGG            &  0.842 & 0.818 \\
${\nu}_6$ & VGG+DA+DO      &  0.884 & 0.872 \\
${\nu}_6$ & Inception+DA+DO&  0.814 & 0.878 \\
${\nu}_6$ & ResNet+DA+DO   &  0.760 & 0.659 \\ 
\hline 
M1:2 & VGG                 & 0.935 & 0.937 \\
M1:2 & VGG+DA+DO           & 0 849 & 0.850 \\
M1:2 & Inception+DA+DO     & 0.882 & 0.890 \\
M1:2 & Resnet+DA+DO+BN     & 0.562 & 0.587 \\
\hline
\end{tabular}
\end{center}
\end{table}

\section{Conclusions}
\label{sec: concl}

The main objective of this paper was to identify which approach, among modern
available $CNN$ models, was i) less prone to overfitting, and ii)
more efficient in predicting the labels of large datasets of asteroids'
resonant images. Using two publically available datasets of
images of asteroids resonant arguments for objects interacting
with the ${\nu}_6$ and M1:2 resonances, and setting aside a validation
set for both databases, we identified three best performing
models for each dataset, corrected (or not) for overfitting
using standard regularization methods such as Data Augmentation,
Dropout, and Batch Normalization, or combinations of such methods.

These best-performing methods were then tested on larger testing datasets,
containing up to 3000 images of asteroidal resonant arguments.
Their performance was measured using the standard {\it accuracy}
metric, and the weighted {\it F-beta} score, a metric apt
to measure the model efficiency for multi-class imbalanced problems,
like the ones studied in this work.
Surprisingly, the simplest method, the {\it VGG}, with or without
regularization, performed best for both cases.  Such models, which
run in seconds on our machine, can now be used to classify thousands
of asteroidal images of resonant arguments with accuracies and F-beta scores
of up to 93.5\%, and they will prove themselves quite valuable when the
Vera C. Rubin surveys will start to discover millions of new asteroids
after the beginning of operations \citep{Jones_2015}.

Obtaining databases for other resonances interacting with asteroidal
populations and performing an optimization study like the one
carried out in this work remain challenges for future research.

\section{Appendix: $CNN$ model performances results}
\label{sec: appendix}

In this appendix, we report plots of cross entropy loss and classification
accuracy for the models tested on the ${\nu}_6$ and M1:2 images databases.
See the discussion in sections~(\ref{sec: nu_6}) and (\ref{sec: M12})
for more information on how to interpret these figures.

\begin{figure*}
  \centering
    \begin{minipage}[c]{0.45\textwidth}
    \centering \includegraphics[width=2.3in]{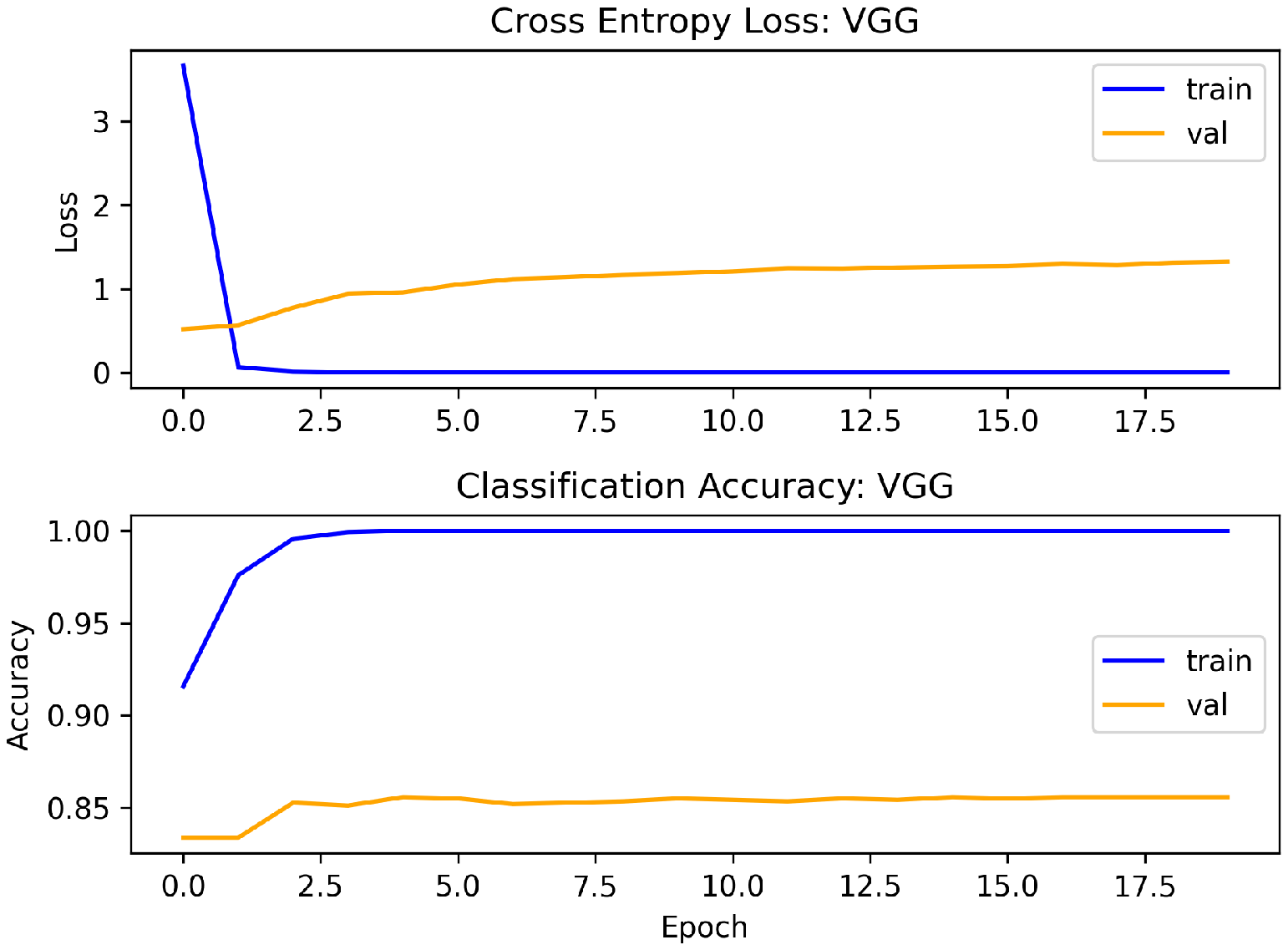}
  \end{minipage}%
  \begin{minipage}[c]{0.45\textwidth}
    \centering \includegraphics[width=2.3in]{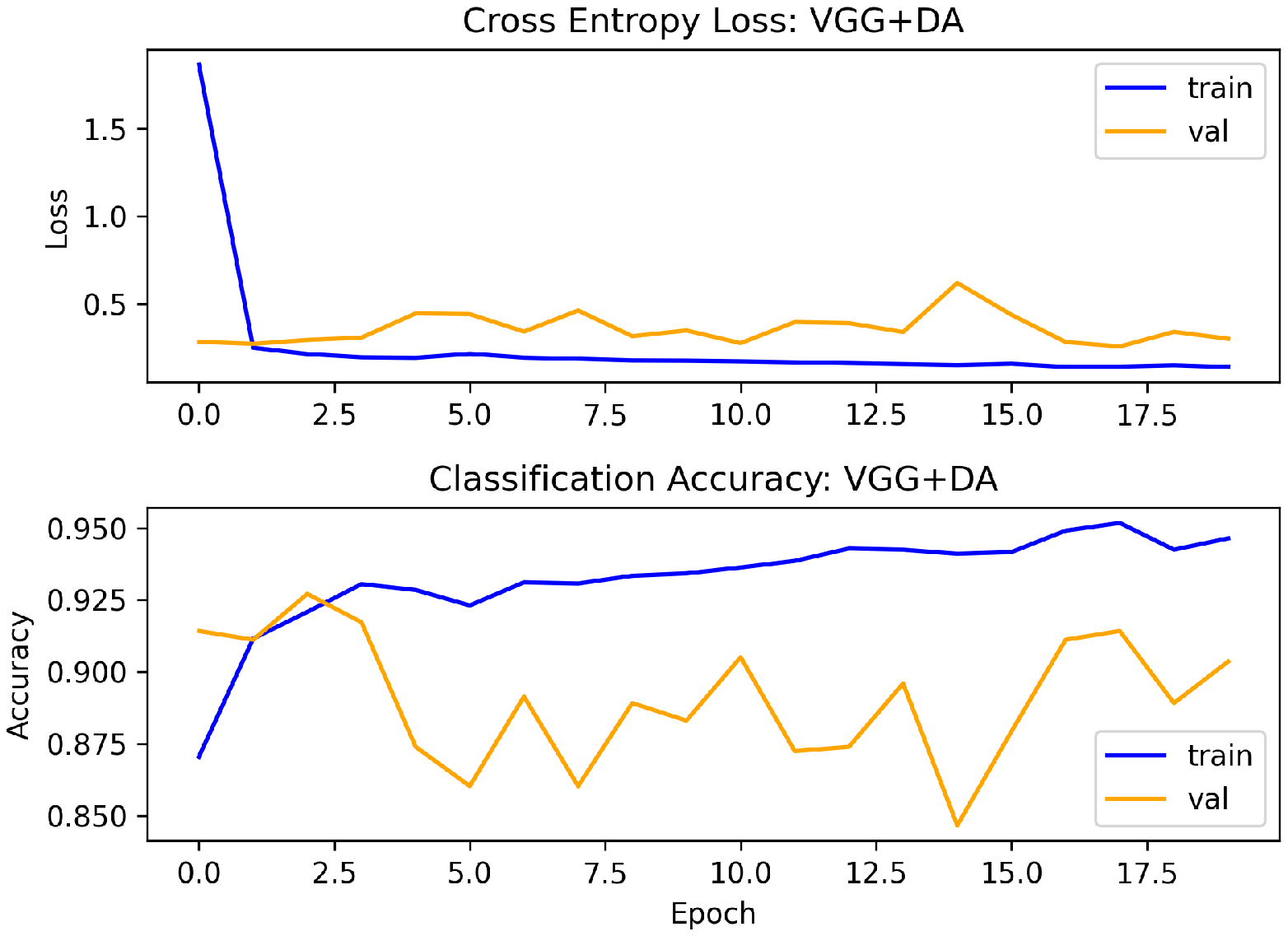}
  \end{minipage}
  \begin{minipage}[c]{0.45\textwidth}
    \centering \includegraphics[width=2.3in]{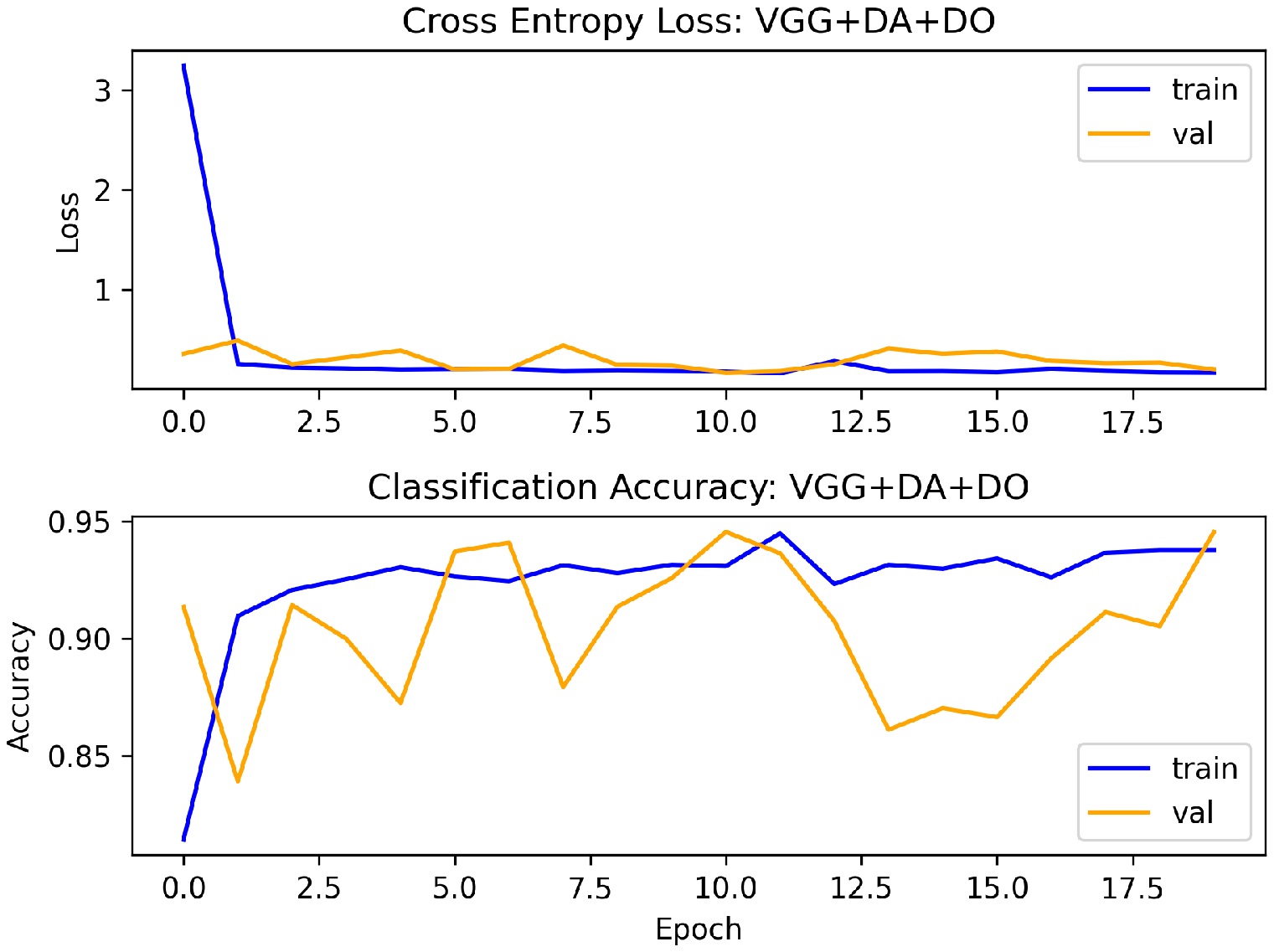}
  \end{minipage}%
    \begin{minipage}[c]{0.45\textwidth}
    \centering \includegraphics[width=2.3in]{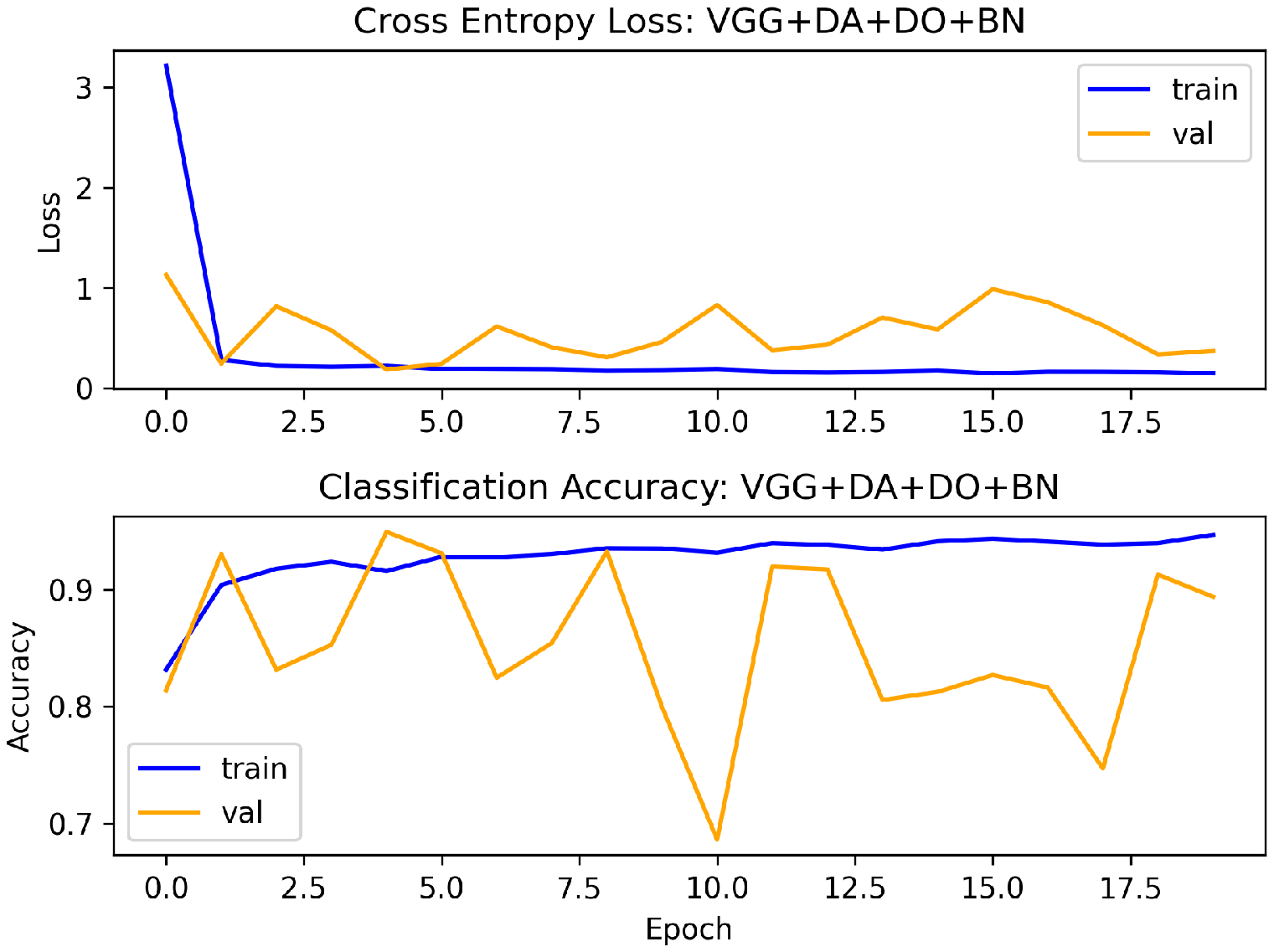}
  \end{minipage}
    \caption{Cross entropy loss and classification accuracy as a function of
      epoch for the VGG model (top left panel), the VGG model plus data
      Augmentation (DA, top right panel), the VGG model plus DA and Dropout
      (DA+DO, bottom left panel), and also including Batch Normalization
      (DA+DO+BN, bottom right panel).  Results obtained for the ${\nu}_6$
      database.}
\label{fig: Metrics_VGG}
\end{figure*}

\begin{figure*}
  \centering
    \begin{minipage}[c]{0.45\textwidth}
    \centering \includegraphics[width=2.3in]{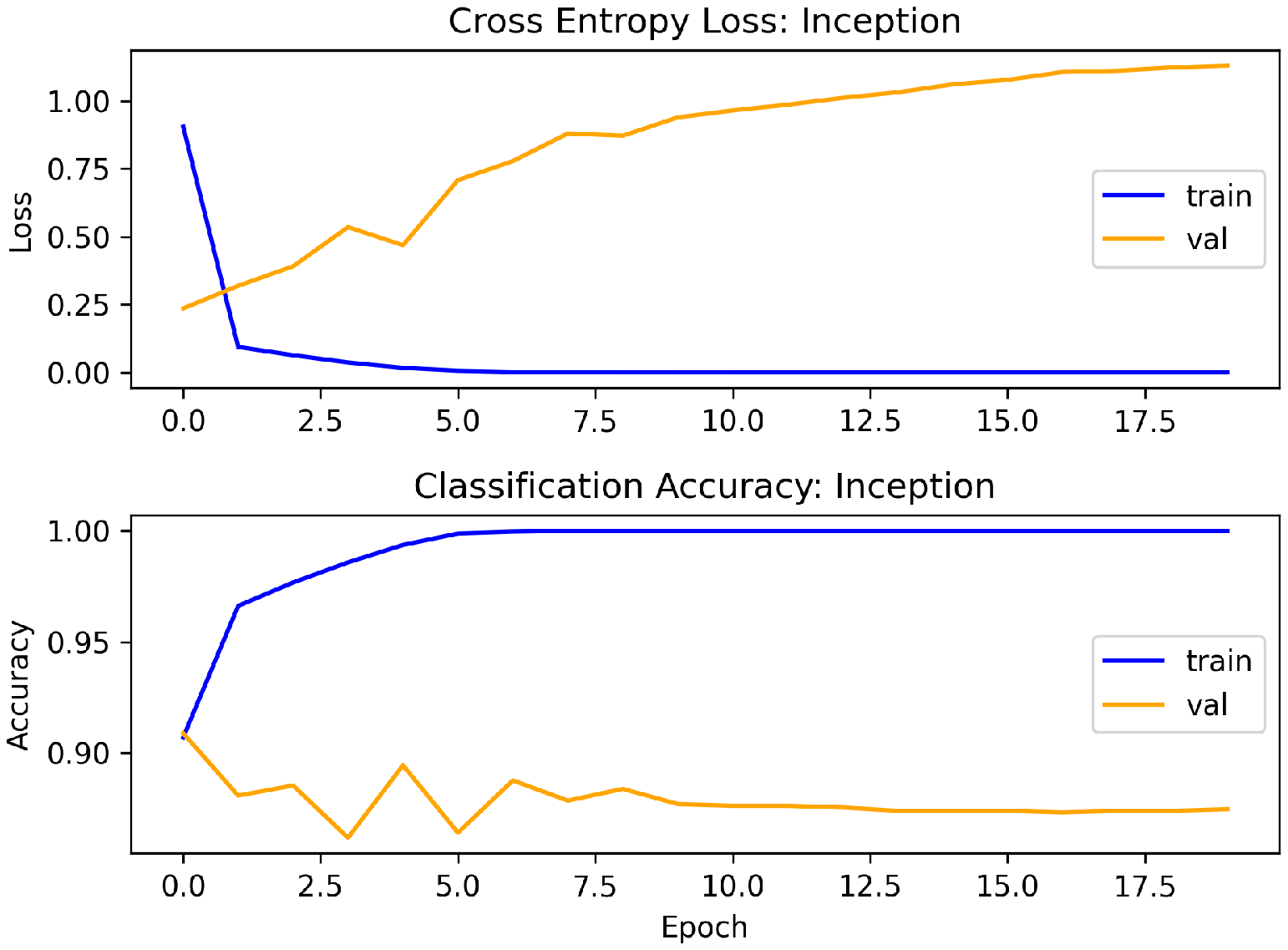}
  \end{minipage}%
  \begin{minipage}[c]{0.45\textwidth}
    \centering \includegraphics[width=2.3in]{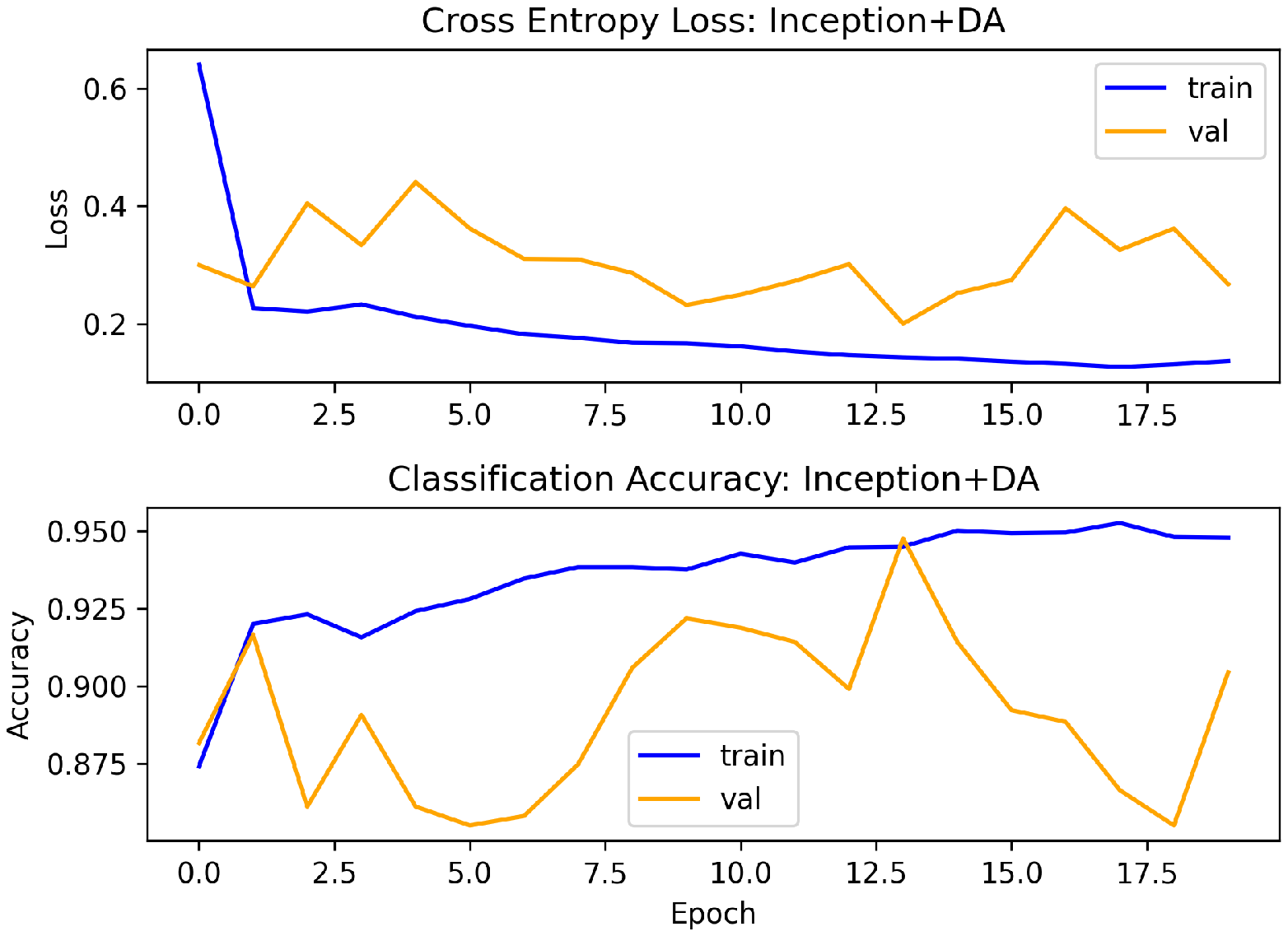}
  \end{minipage}
  \begin{minipage}[c]{0.45\textwidth}
    \centering \includegraphics[width=2.3in]{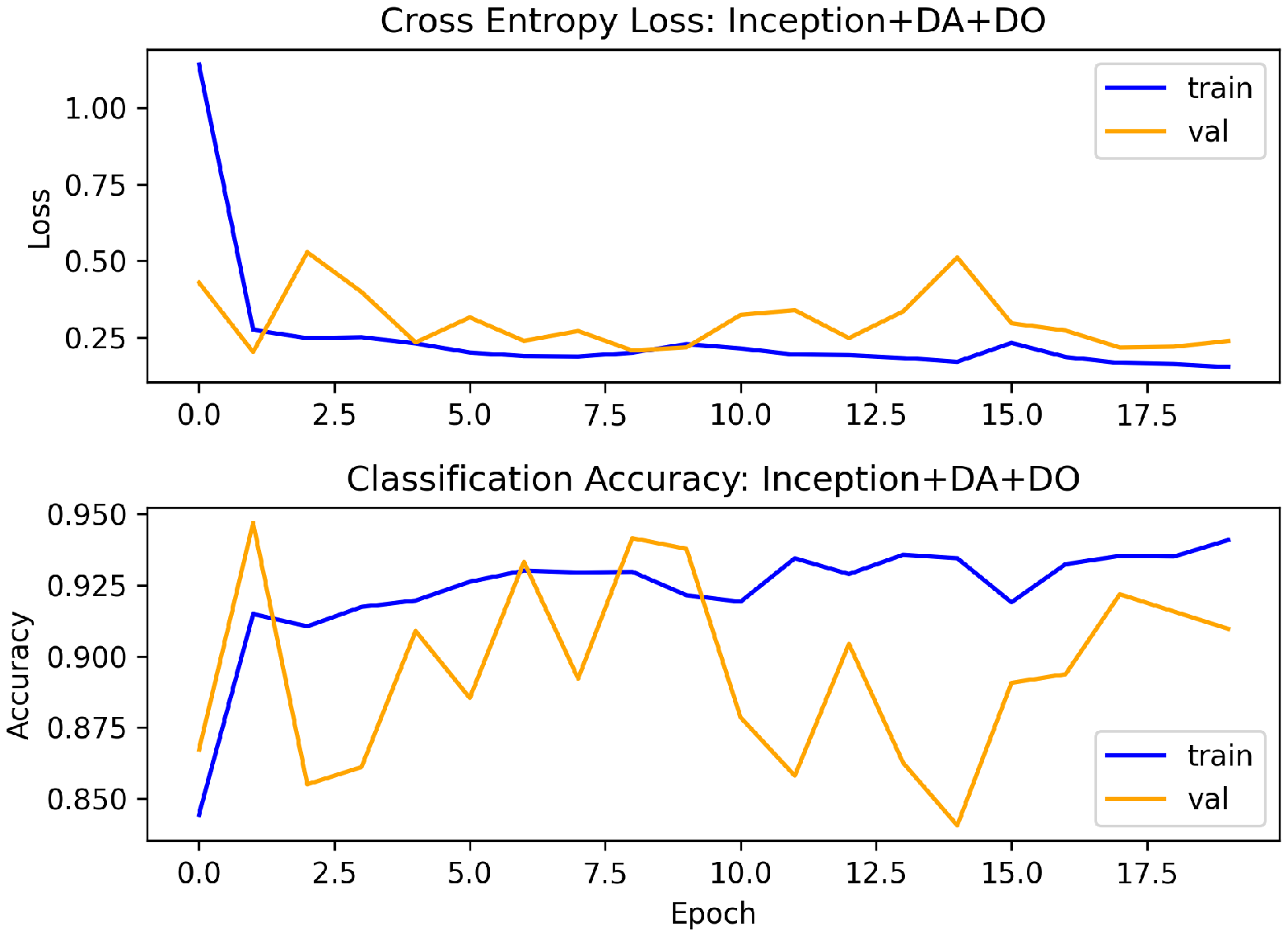}
  \end{minipage}%
    \begin{minipage}[c]{0.45\textwidth}
    \centering \includegraphics[width=2.3in]{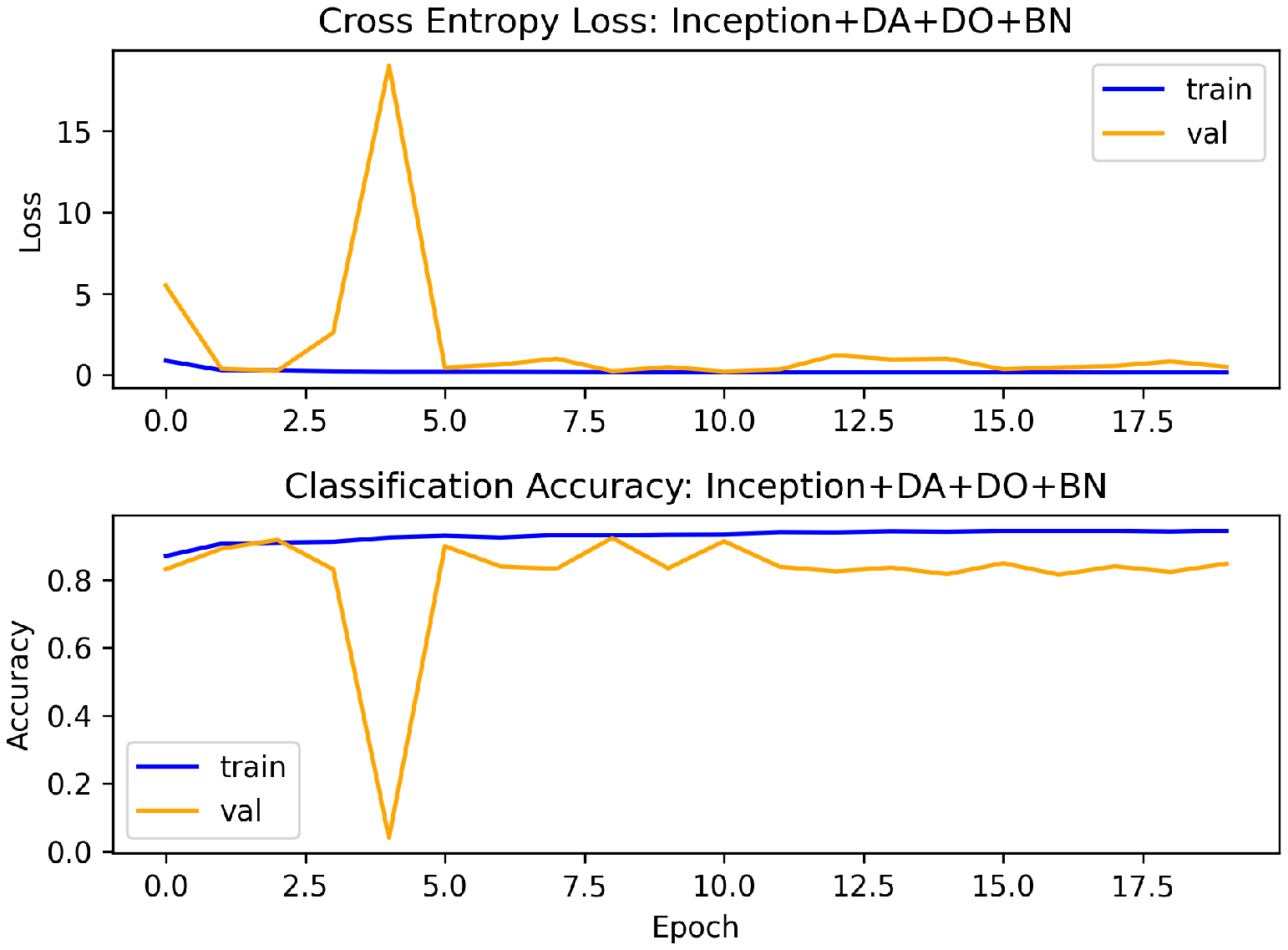}
  \end{minipage}
    \caption{The same as in figure~(\ref{fig: Metrics_VGG}), but for the
      outcome of the Inception models.}
\label{fig: Metrics Inception}
\end{figure*}

\begin{figure*}
  \centering
    \begin{minipage}[c]{0.45\textwidth}
    \centering \includegraphics[width=2.3in]{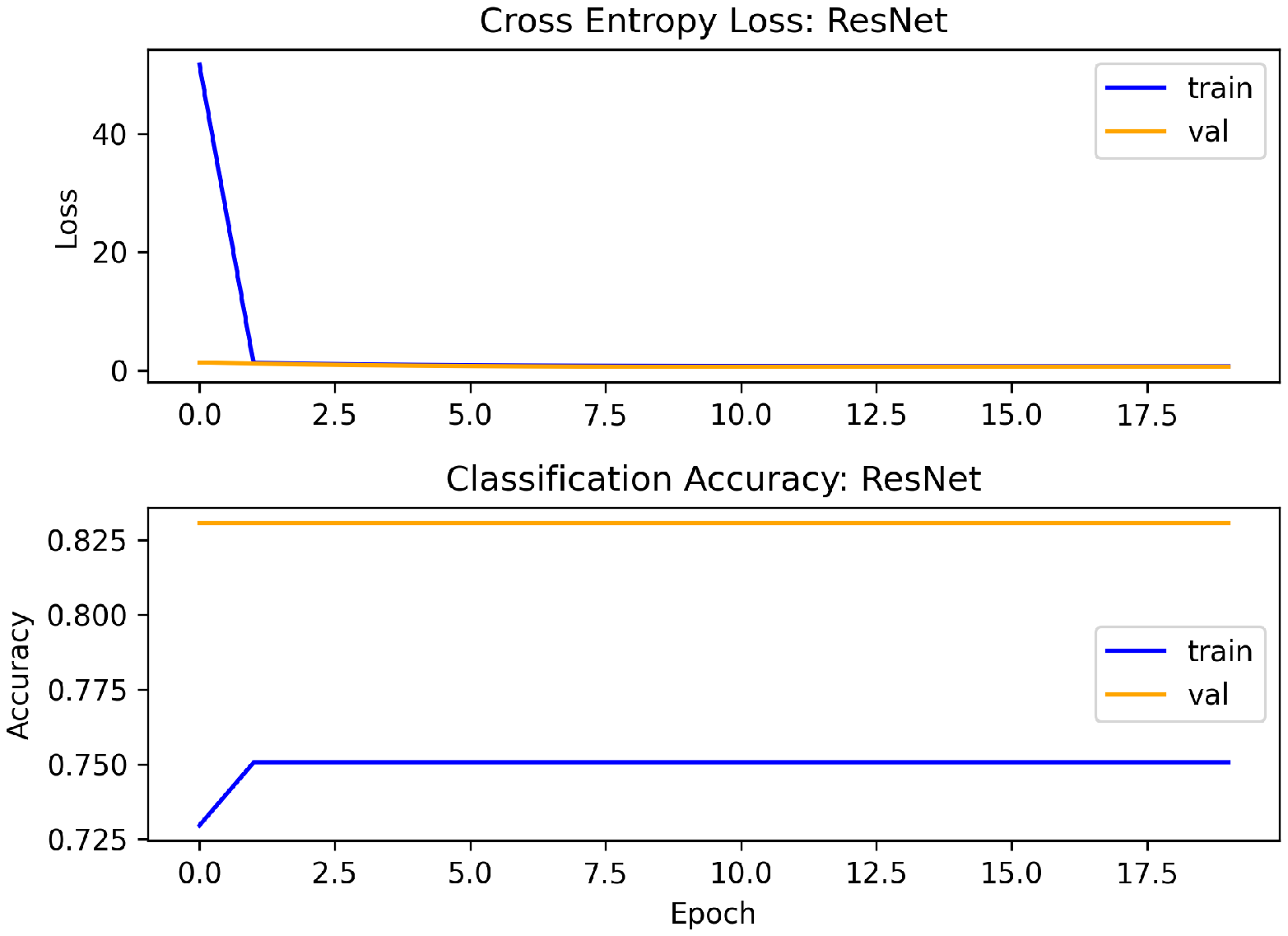}
  \end{minipage}%
  \begin{minipage}[c]{0.45\textwidth}
    \centering \includegraphics[width=2.3in]{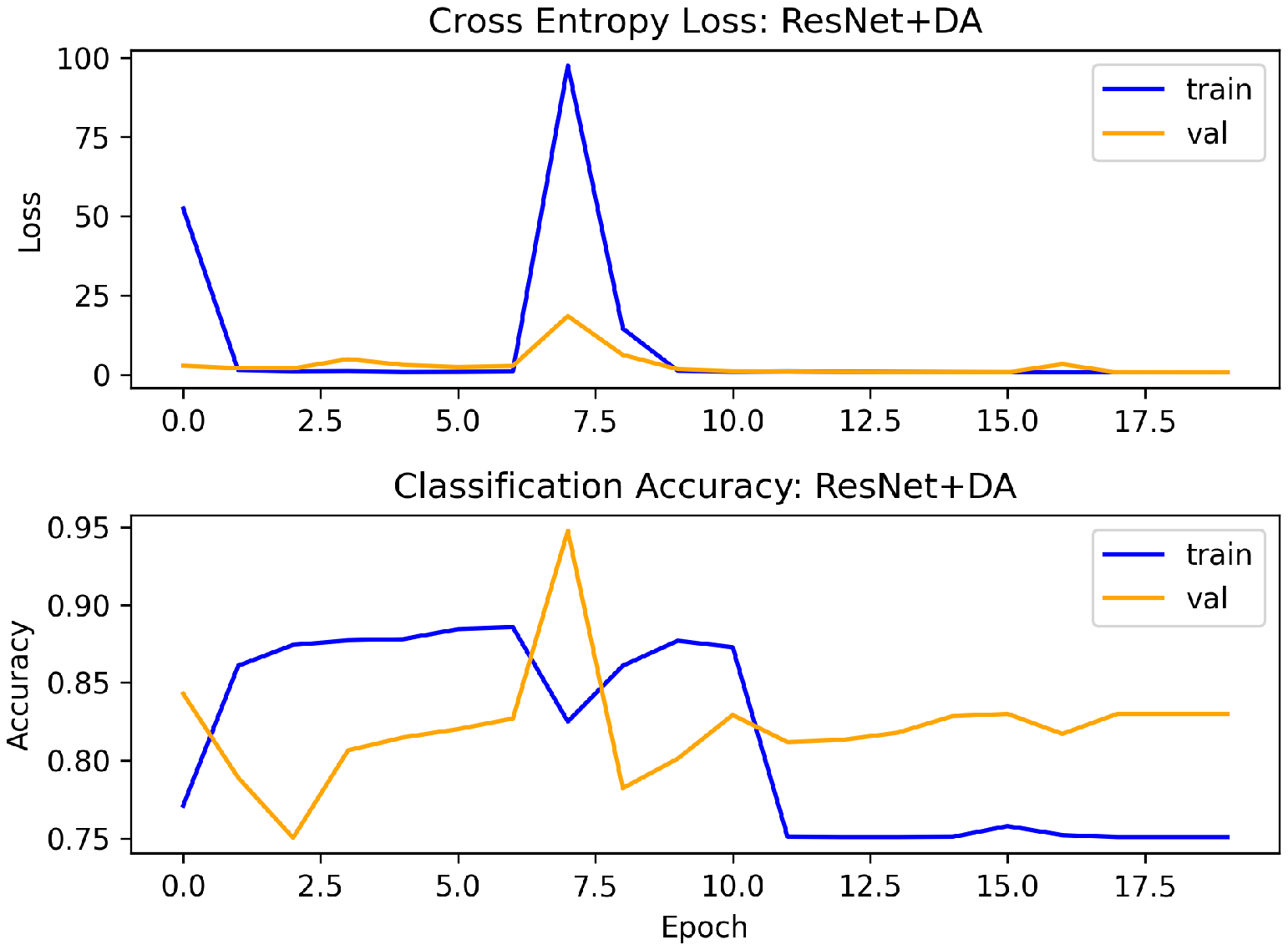}
  \end{minipage}
  \begin{minipage}[c]{0.45\textwidth}
    \centering \includegraphics[width=2.3in]{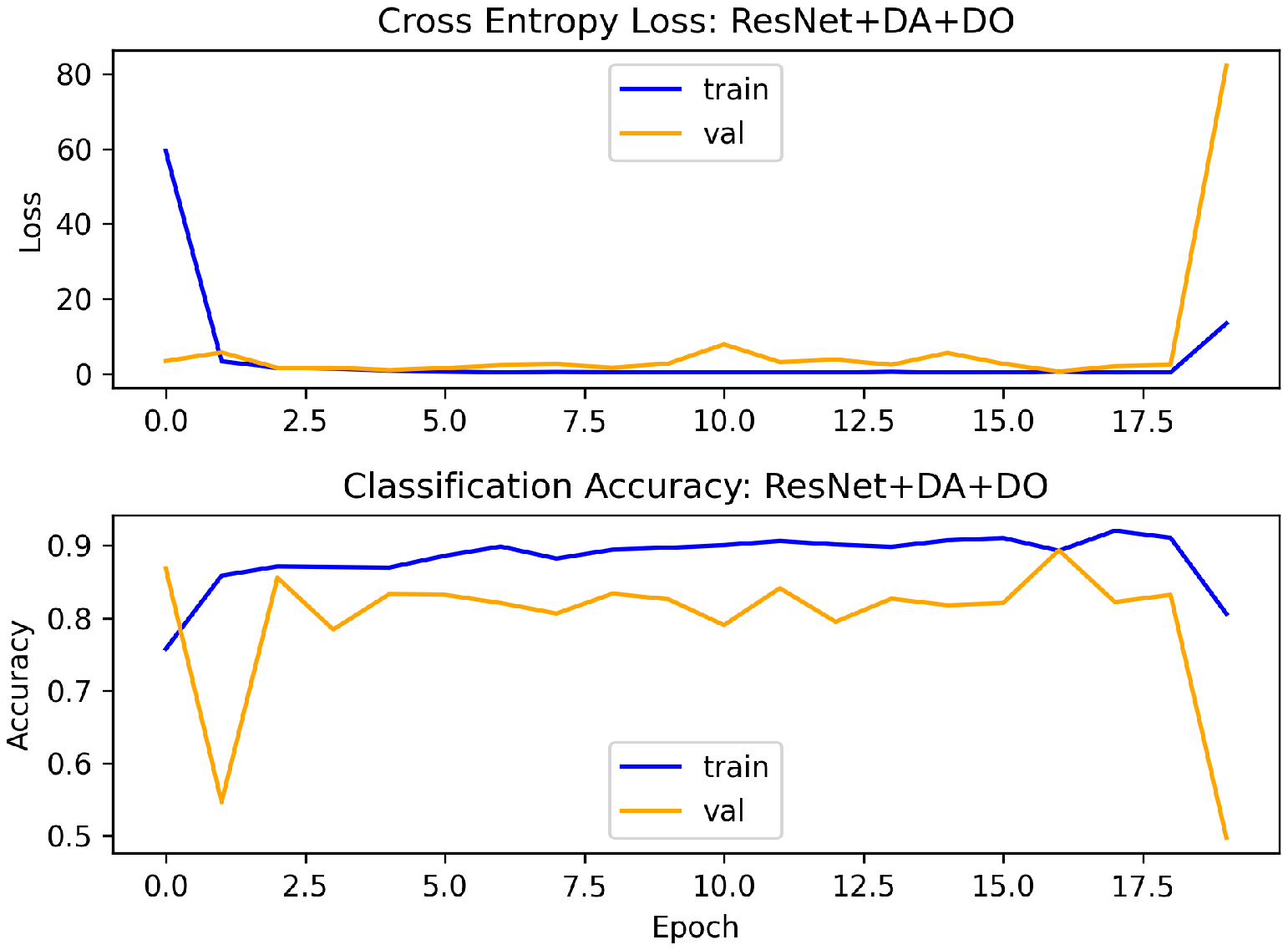}
  \end{minipage}%
    \begin{minipage}[c]{0.45\textwidth}
    \centering \includegraphics[width=2.3in]{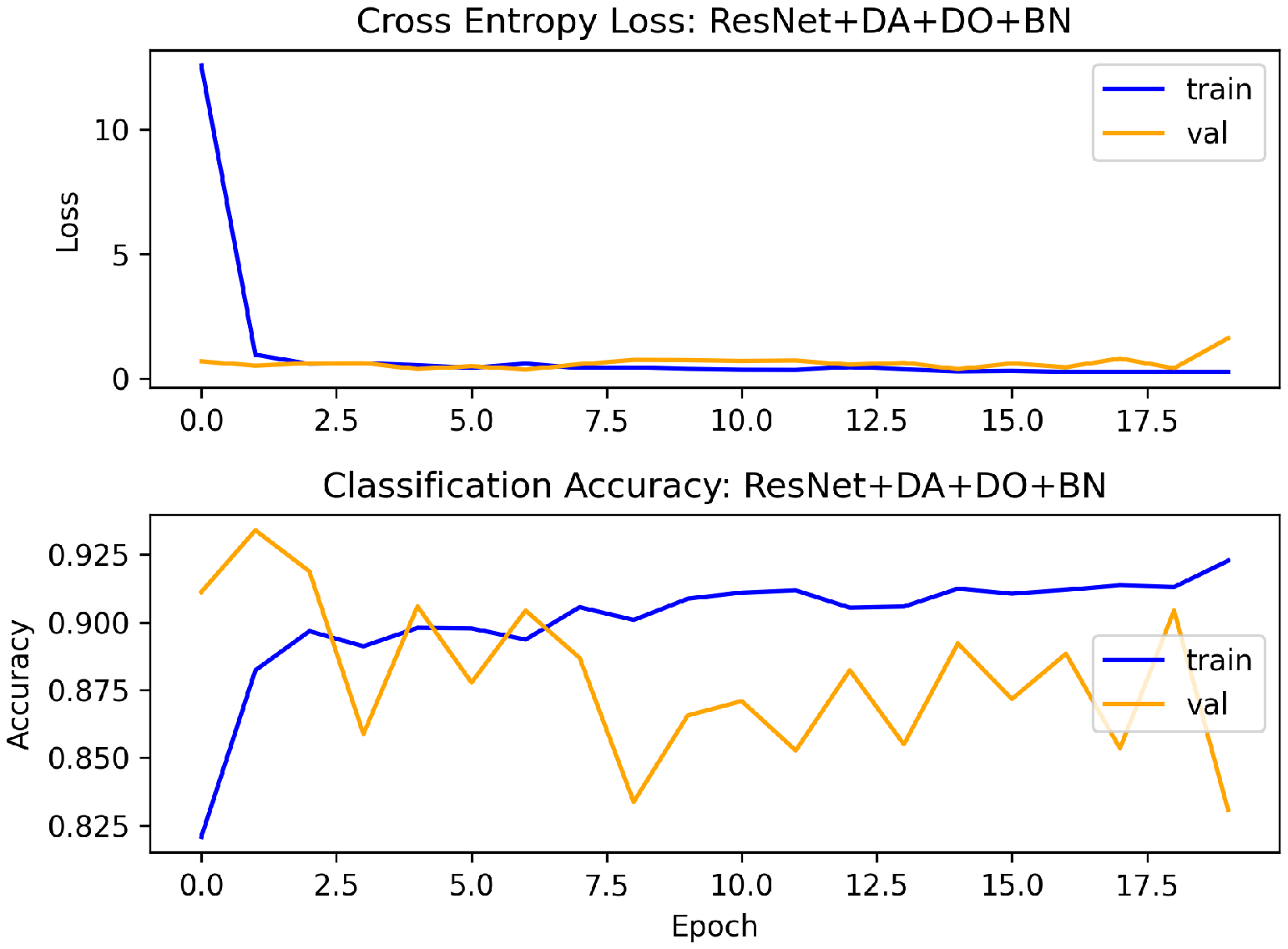}
  \end{minipage}
    \caption{The same as in figure~(\ref{fig: Metrics_VGG}), but for the
      outcome of the ResNet models.}
\label{fig: Metrics ResNet}
\end{figure*}

\begin{figure*}
  \centering
    \begin{minipage}[c]{0.45\textwidth}
    \centering \includegraphics[width=2.3in]{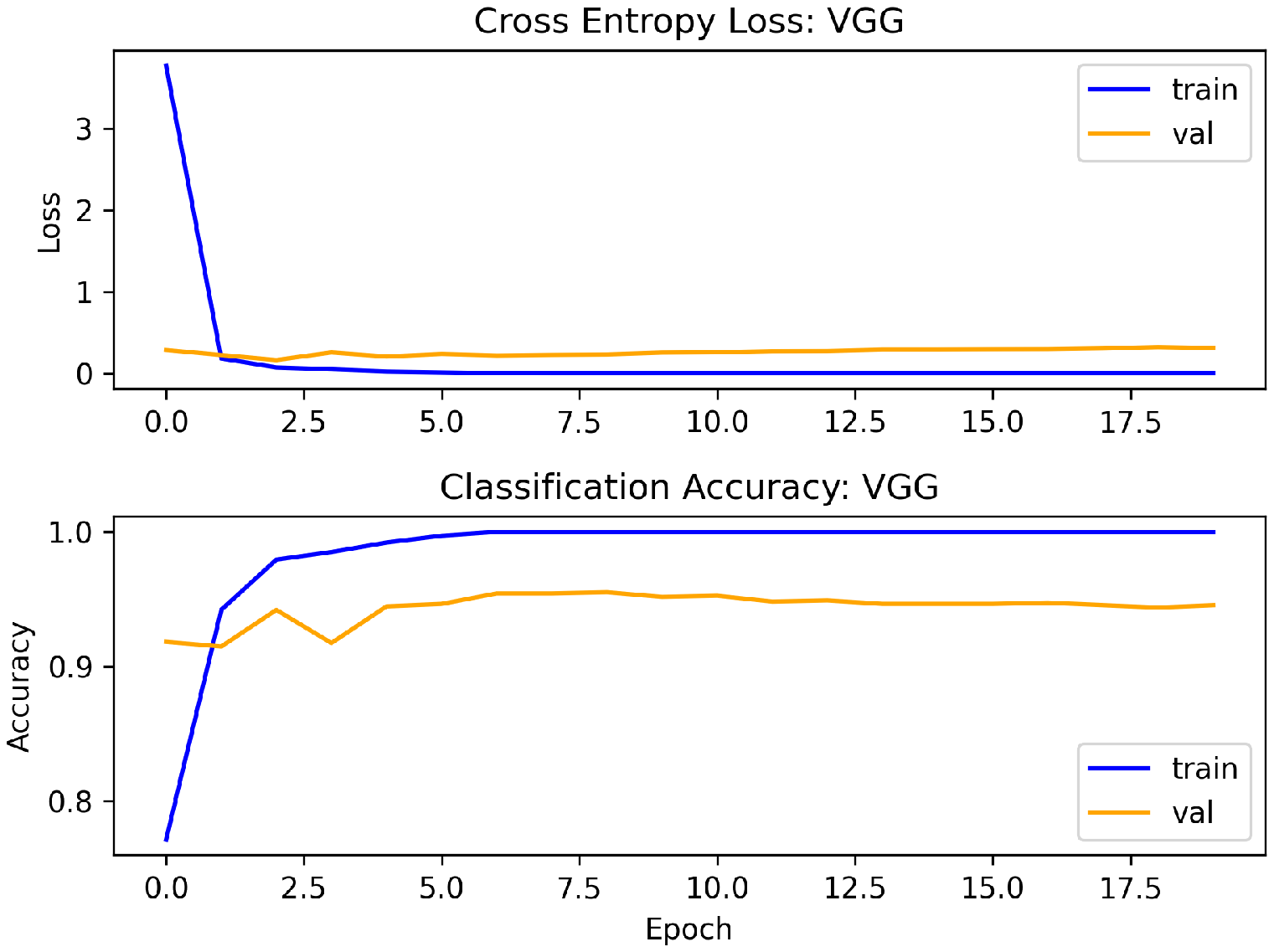}
  \end{minipage}%
  \begin{minipage}[c]{0.45\textwidth}
    \centering \includegraphics[width=2.3in]{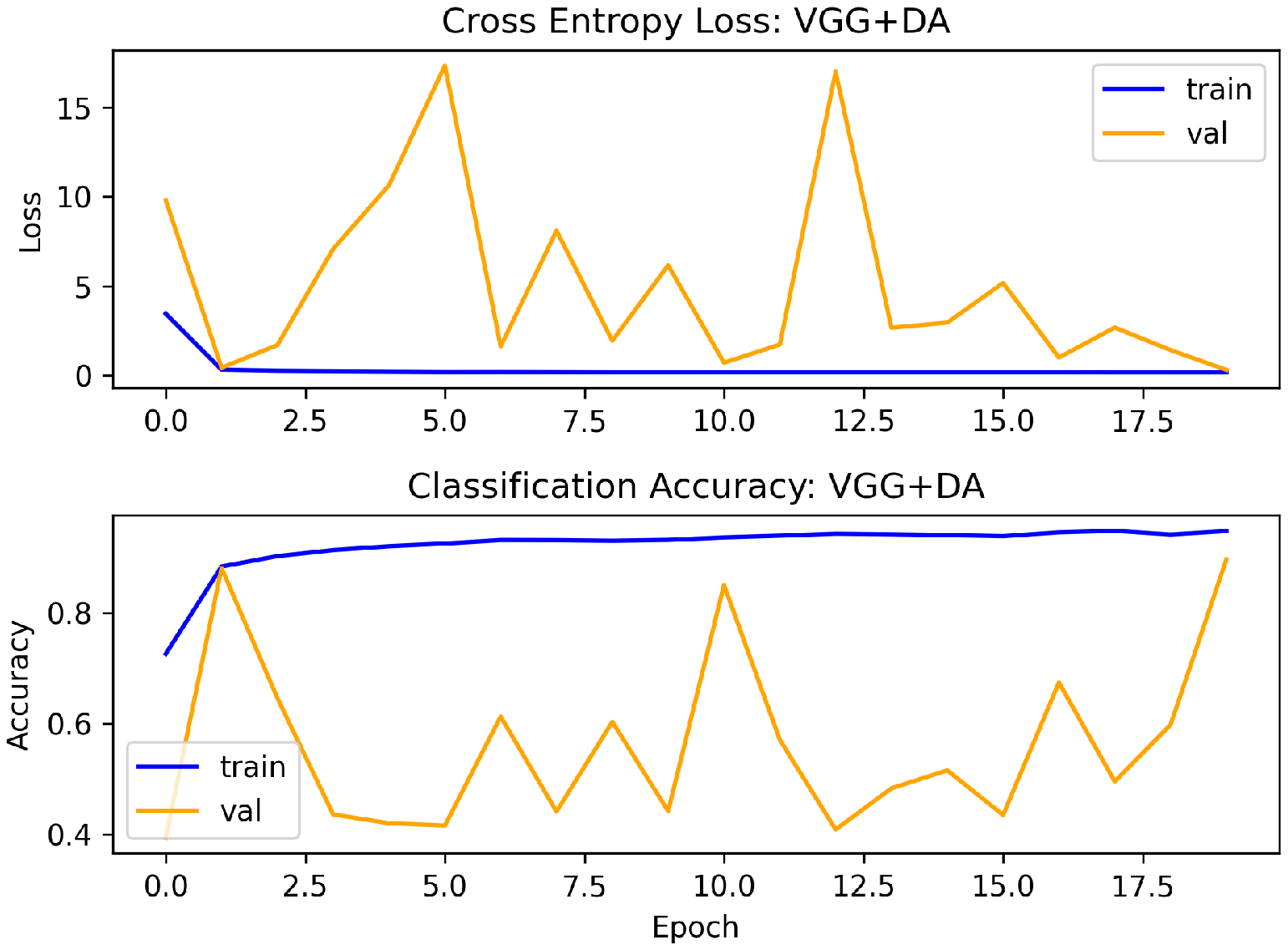}
  \end{minipage}
  \begin{minipage}[c]{0.45\textwidth}
    \centering \includegraphics[width=2.3in]{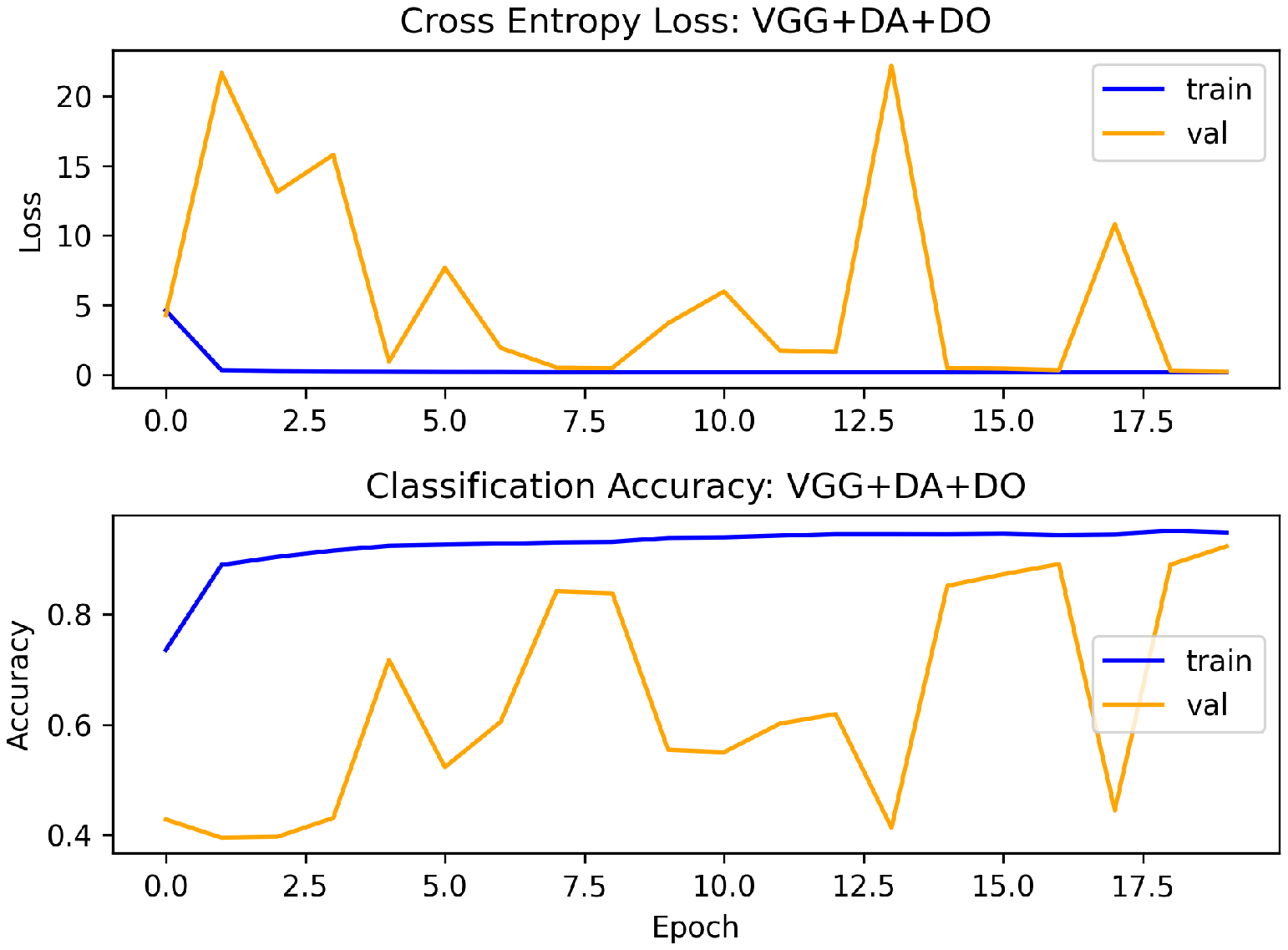}
  \end{minipage}%
    \begin{minipage}[c]{0.45\textwidth}
    \centering \includegraphics[width=2.3in]{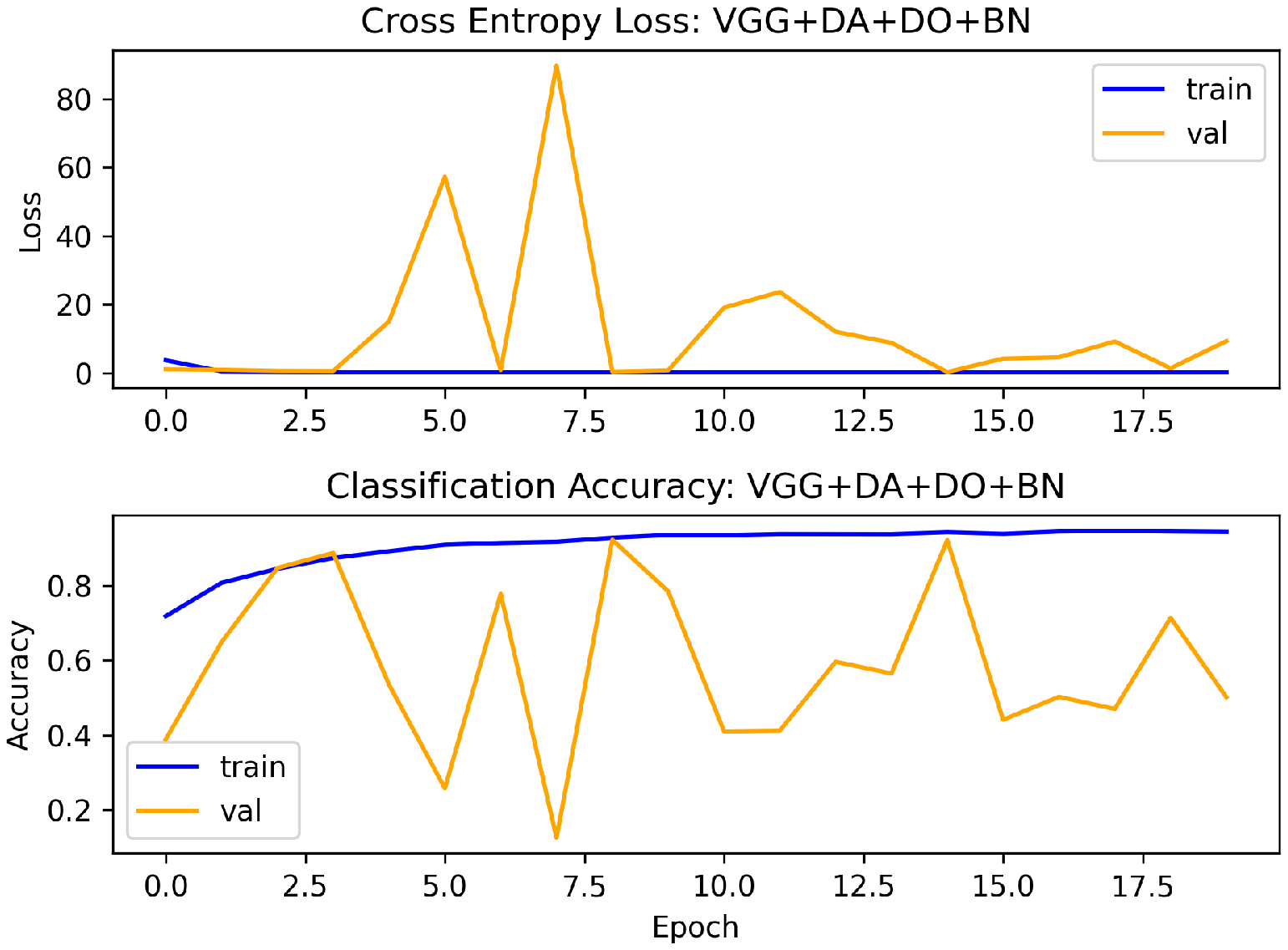}
  \end{minipage}
    \caption{Cross entropy loss and classification accuracy as a function of
      epoch for the VGG model (top left panel), the VGG model plus data
      Augmentation (DA, top right panel), the VGG model plus DA and Dropout
      (DA+DO, bottom left panel), and also including Batch Normalization
      (DA+DO+BN, bottom right panel).  Results obtained for the M1:2
      database.}
\label{fig: Metrics_VGG_M12}
\end{figure*}

\begin{figure*}
  \centering
    \begin{minipage}[c]{0.45\textwidth}
    \centering \includegraphics[width=2.3in]{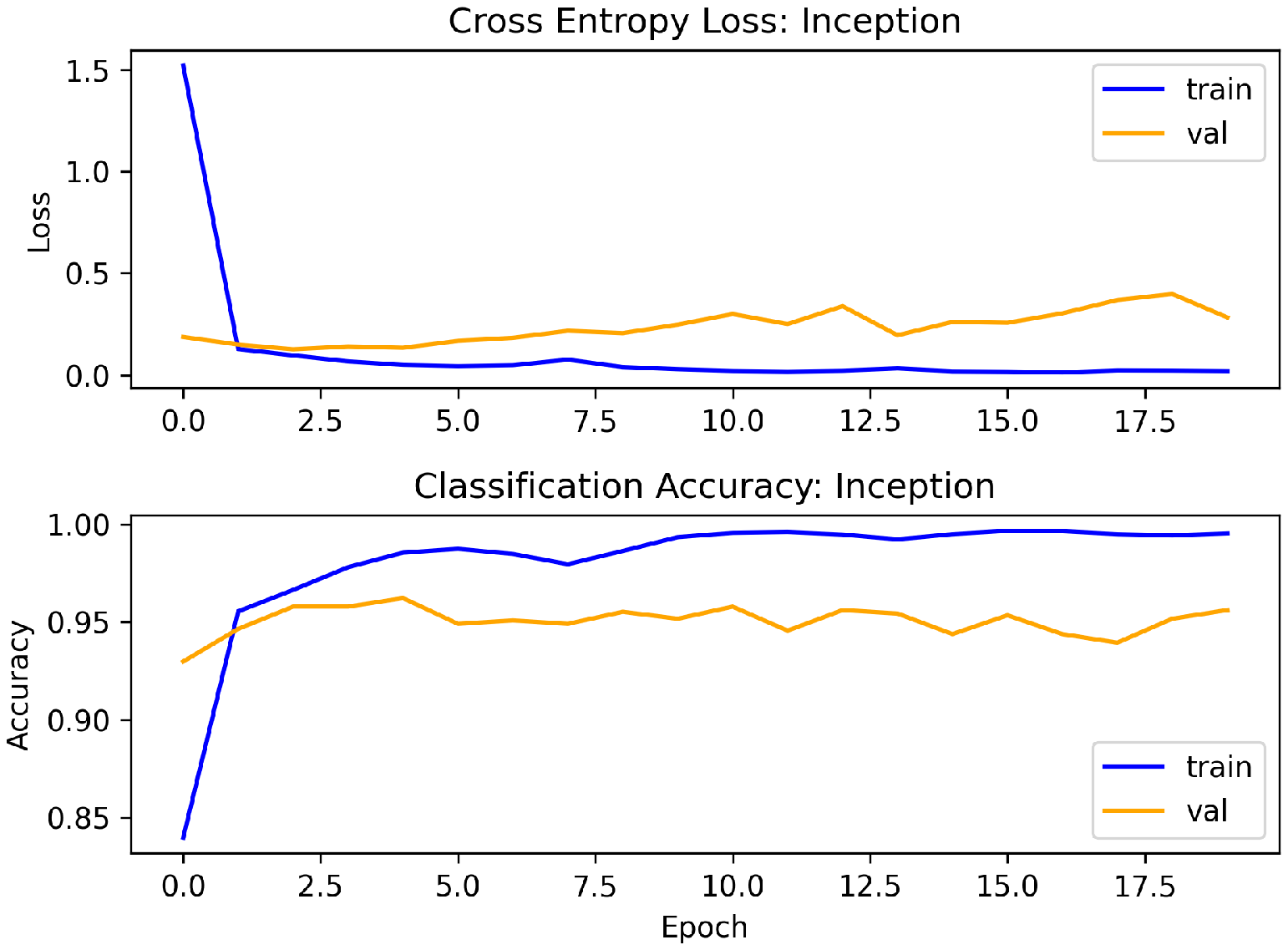}
  \end{minipage}%
  \begin{minipage}[c]{0.45\textwidth}
    \centering \includegraphics[width=2.3in]{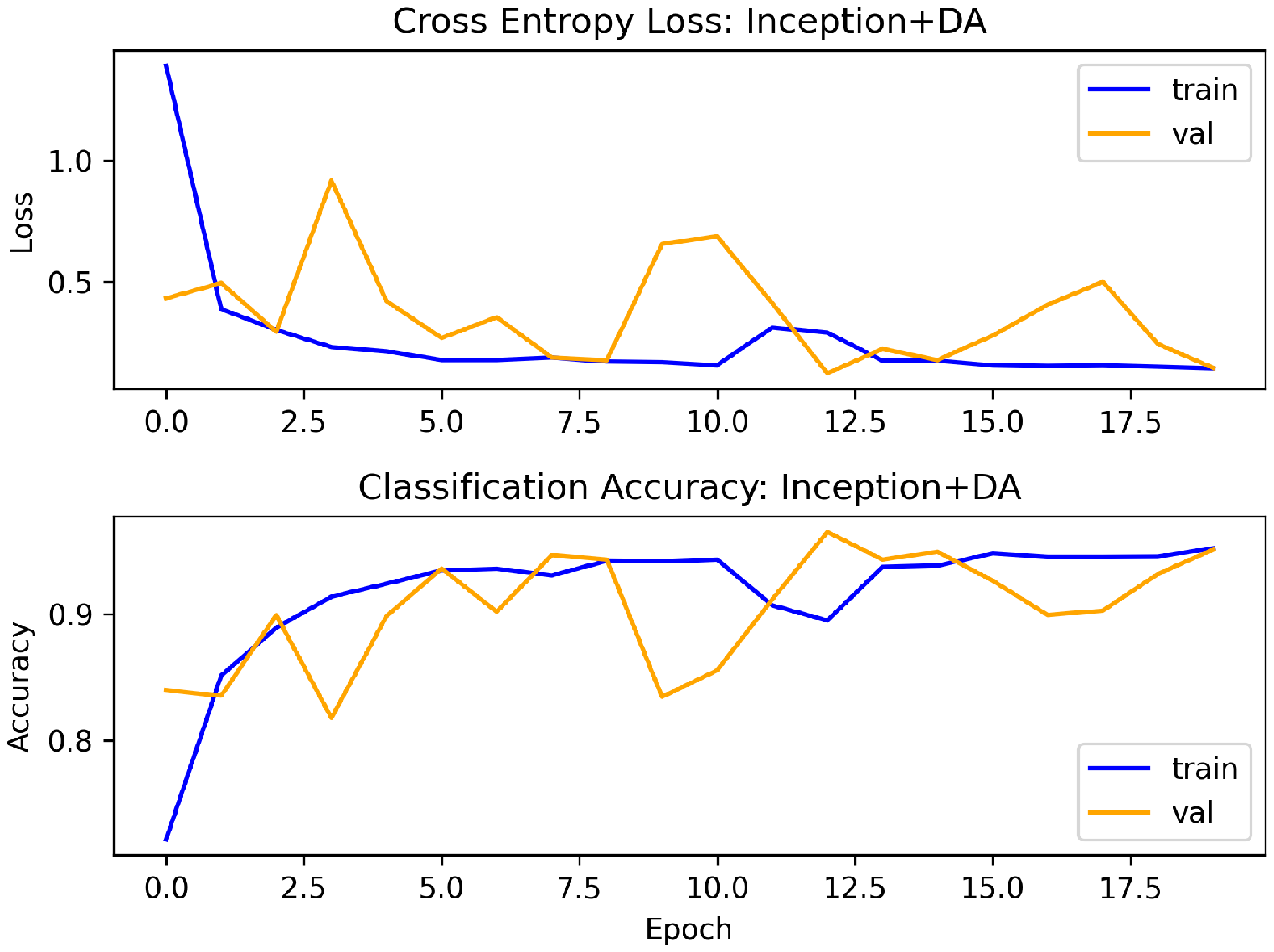}
  \end{minipage}
  \begin{minipage}[c]{0.45\textwidth}
    \centering \includegraphics[width=2.3in]{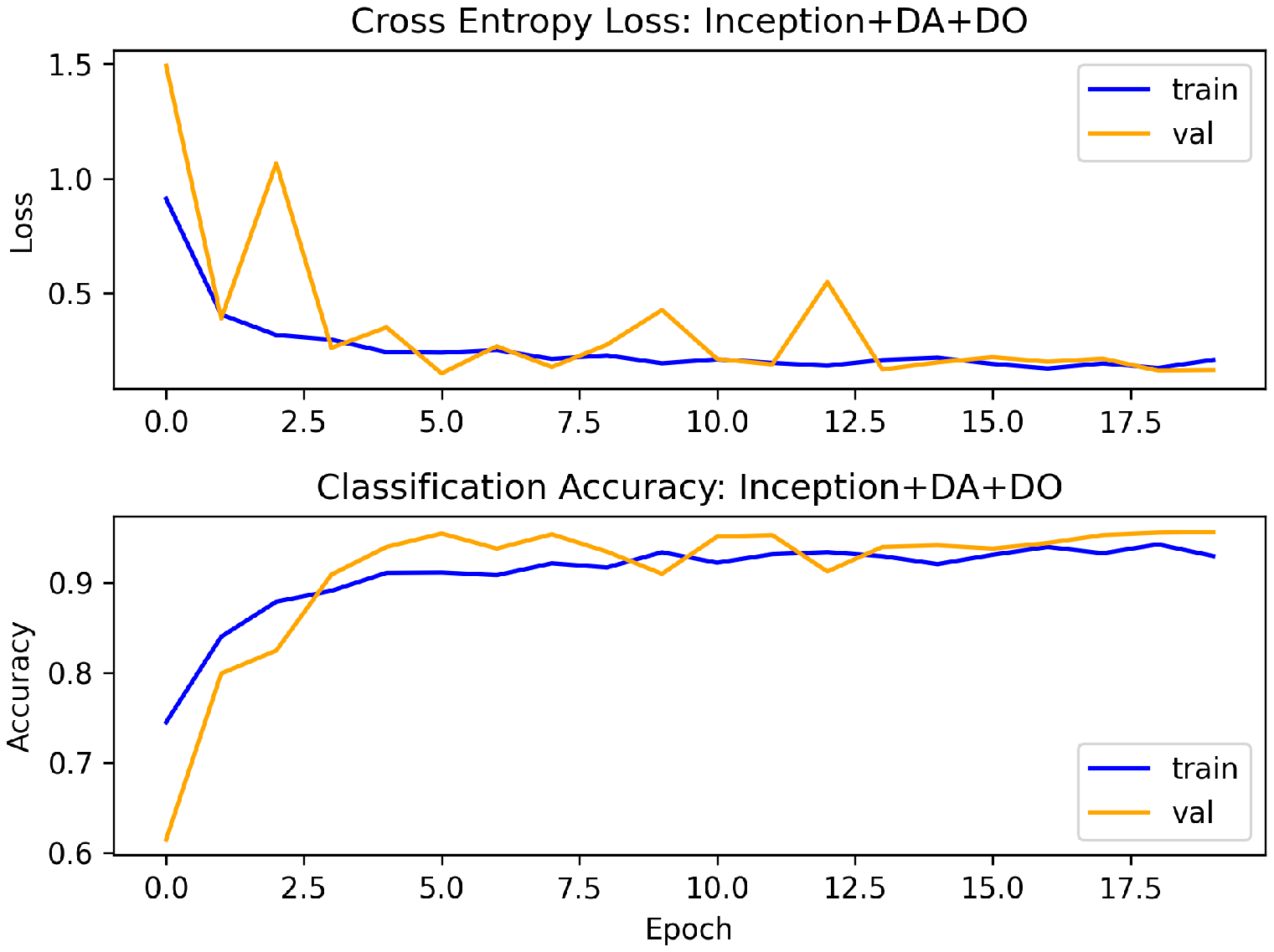}
  \end{minipage}%
    \begin{minipage}[c]{0.45\textwidth}
    \centering \includegraphics[width=2.3in]{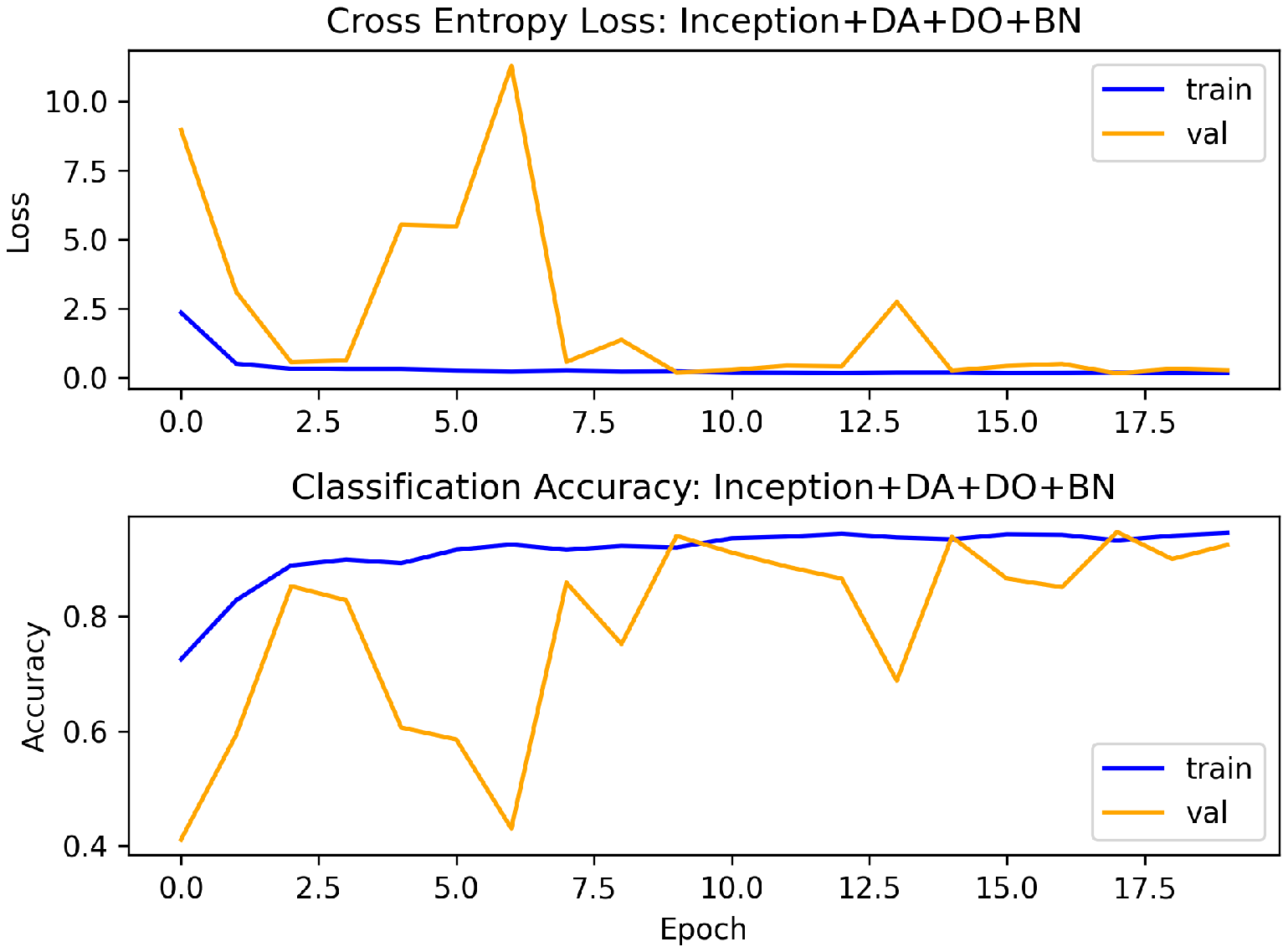}
  \end{minipage}
    \caption{The same as in figure~(\ref{fig: Metrics_VGG_M12}), but for the
      outcome of the Inception models.}
\label{fig: Metrics Inception_M12}
\end{figure*}

\begin{figure*}
  \centering
    \begin{minipage}[c]{0.45\textwidth}
    \centering \includegraphics[width=2.3in]{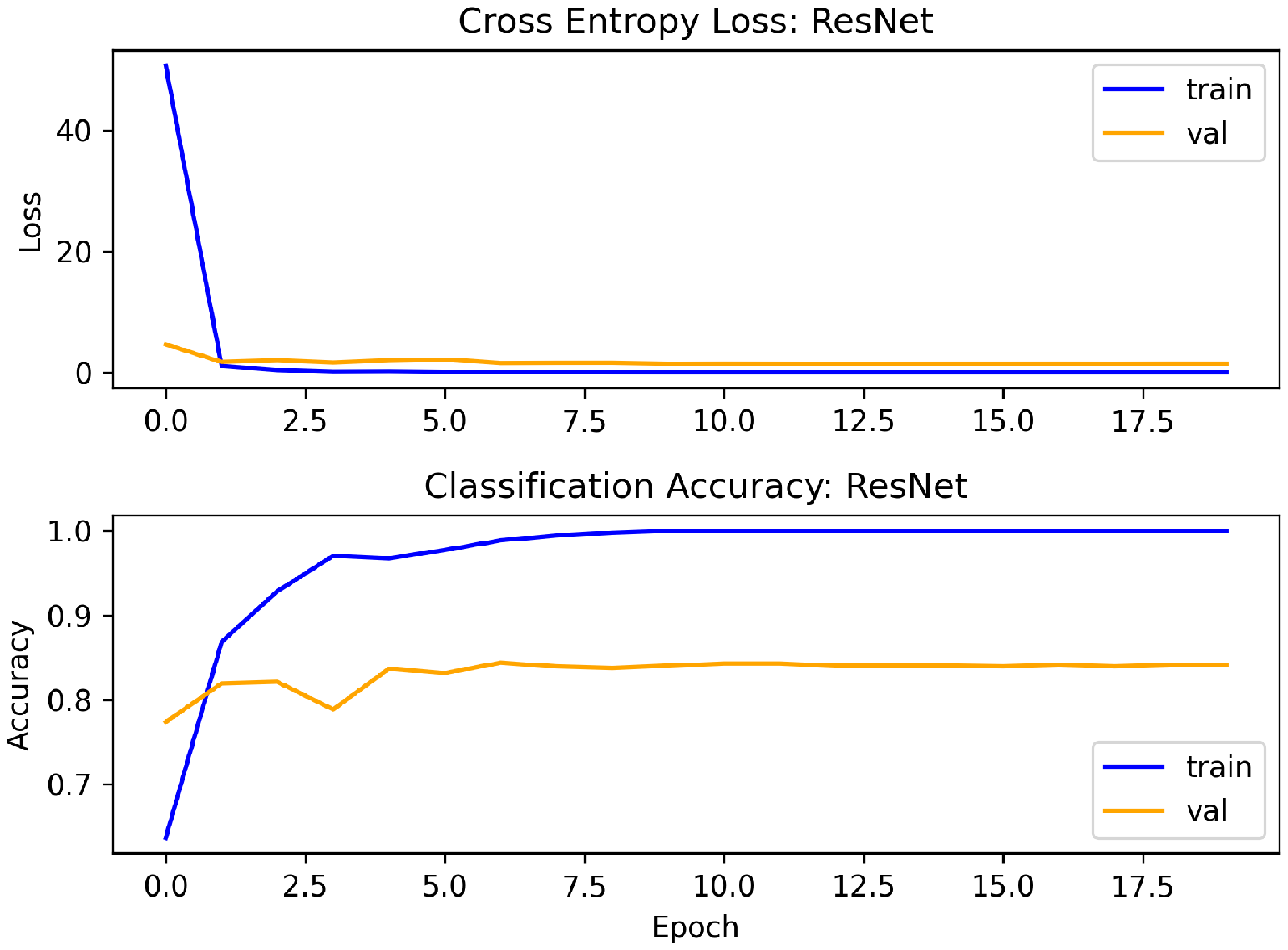}
  \end{minipage}%
  \begin{minipage}[c]{0.45\textwidth}
    \centering \includegraphics[width=2.3in]{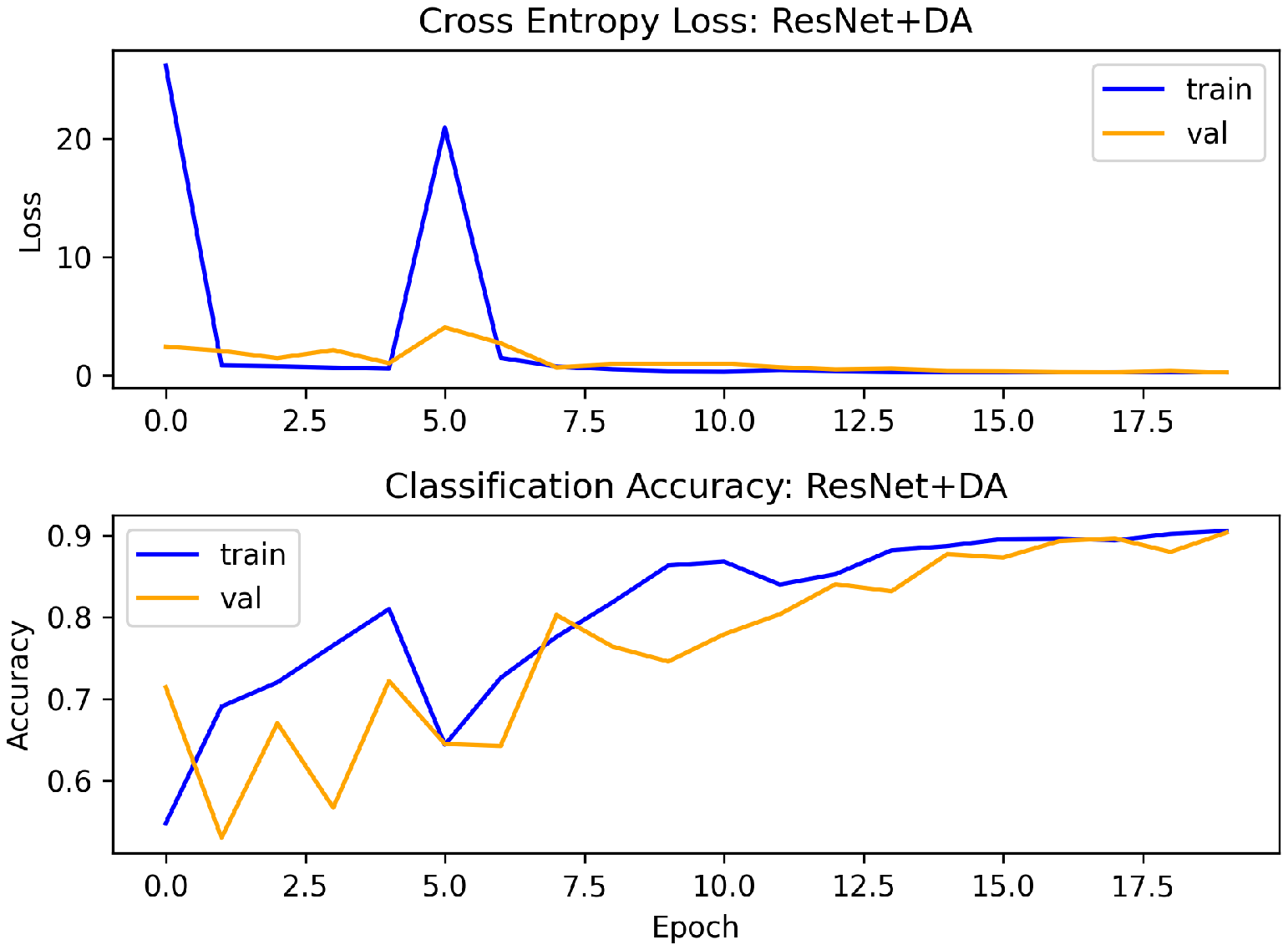}
  \end{minipage}
  \begin{minipage}[c]{0.45\textwidth}
    \centering \includegraphics[width=2.3in]{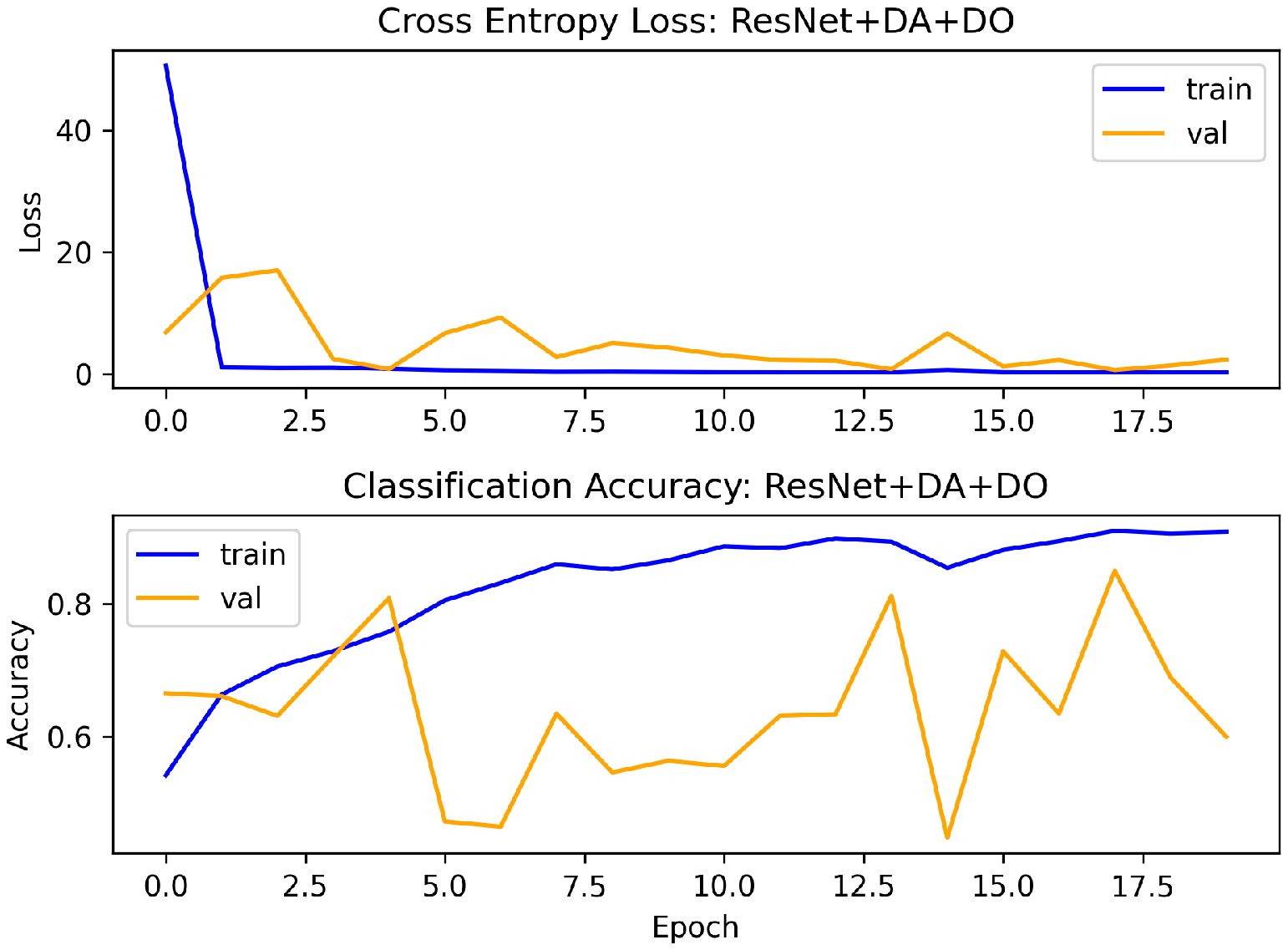}
  \end{minipage}%
    \begin{minipage}[c]{0.45\textwidth}
    \centering \includegraphics[width=2.3in]{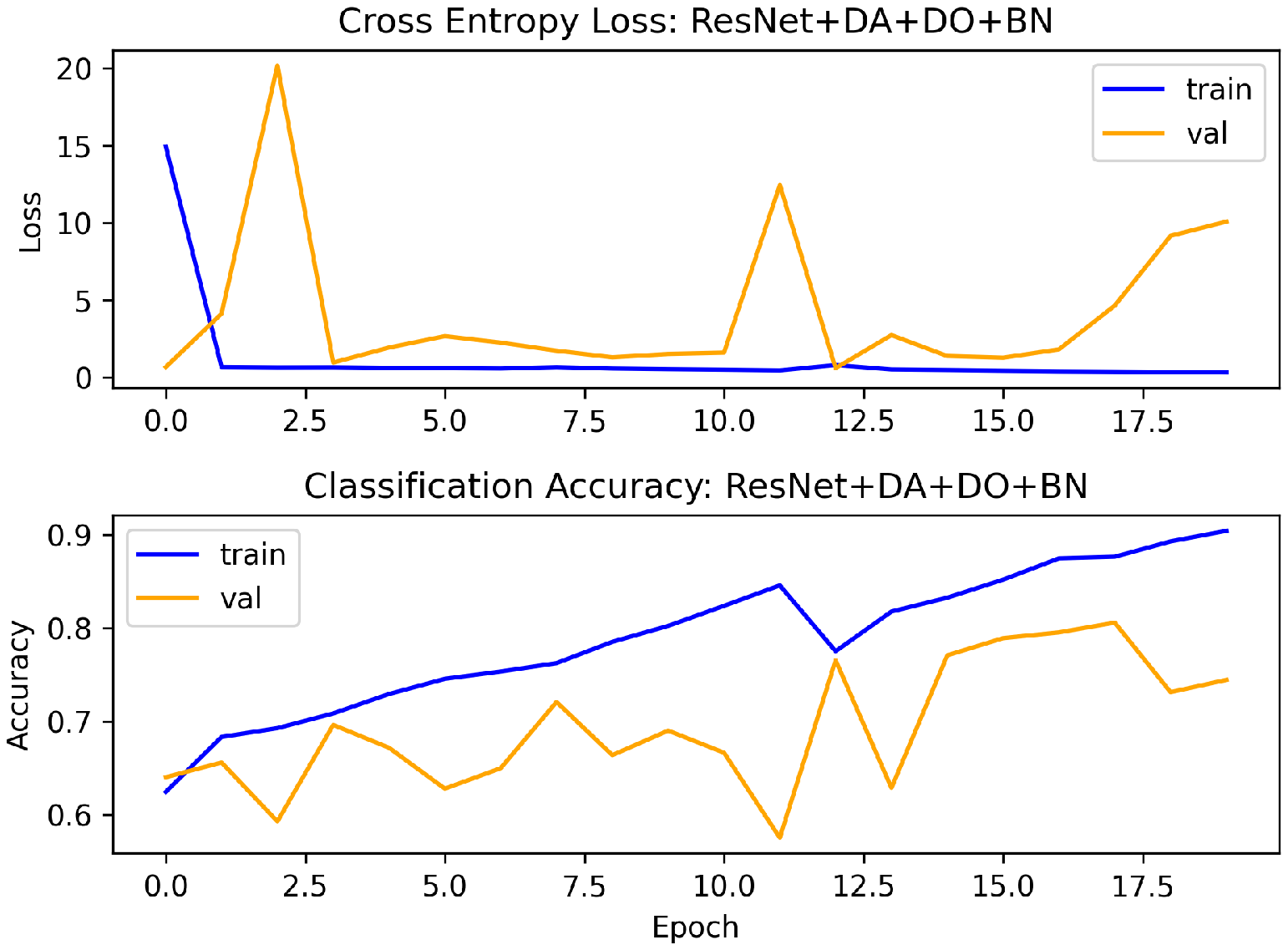}
  \end{minipage}
    \caption{The same as in figure~(\ref{fig: Metrics_VGG_M12}), but for the
      outcome of the ResNet models.}
\label{fig: Metrics ResNet_M12}
\end{figure*}

\section*{Acknowledgments}

We are grateful to two anonymous reviewers for helpful comments
and suggestions.  We would like to thank the Brazilian National Research Council
(CNPq, grant 304168/2021-1), the São Paulo Research Foundation (FAPESP,
grant 2016/024561-0), and the ``Coorde\-na\c{c}\~{a}o de Aperfei\c{c}oamento
de Pessoal de N\'{i}vel Superior'' (CAPES, grant 88887.675709/2022-00). This
is a publication from the MASB (Machine-learning applied to small bodies, \\
https://\-valeriocarruba.github.\-io/Site-MASB/) research group.  Questions
on this paper can also be sent to the group email address: \\
{\it mlasb2021@gmail.com}.

\section*{Conflict of interest}
The authors declare that they have no conflict of interest.

\section{Author contributions}

All authors contributed to the study conception and design. Material
preparation, and data collection were performed by Valerio Carruba, and
Safwan Aljbaae. The first draft of the manuscript was written by Valerio
Carruba and all authors commented on previous versions of the manuscript.
All authors read and approved the final manuscript.

\section{Data availability}

The images database for the ${\nu}_6$ resonance was published in
\citet{2022MNRAS.514.4803C} and is available at this link:

\vspace{0.1cm}
https://drive.google.com/drive/folders/\\
\-1oxdXquibdYYP965AgzoMjFop-ZZXcHni
\vspace{0.1cm}

With the exception of the images used for the testing set in section
~(\ref{sec: larger_data}), the database for the M1:2 resonance
was published in \citet{2021MNRAS.504..692C} and is available at:

\vspace{0.1cm}
https://drive.google.com/file/d/\\
\-1RsDoMh8iMwZhD-fnkYSs9hiWmg96SZf0/\\
\-view?usp=sharing\\
\vspace{0.1cm}

The testing set of images for M1:2 resonance used in
section~(\ref{sec: larger_data}) can be obtained from the first author upon
reasonable request.

\section{Code availability}

The codes developed for this work are publically available at this
GitHub repository:

\vspace{0.1cm}
https://github.com/valeriocarruba/CNN-Optimization
\vspace{0.1cm}

\bibliographystyle{spbasic}      
\bibliography{mybib}

\label{lastpage}

\end{document}